\documentclass[aps,prx, amsmath,amssymb,print]{revtex4-2}

\usepackage{graphicx}
\usepackage{dcolumn}
\usepackage{bm}
\usepackage{enumerate}
\usepackage{lipsum, babel}
\usepackage{xcolor}
\usepackage{hyperref}
\usepackage[title]{appendix}

\usepackage{lineno}
\usepackage{mathtools}
\usepackage[normalem]{ulem}

\def\be{\begin{equation}}
\def\ee{\end{equation}}
\def\bea{\begin{eqnarray}}
\def\eea{\end{eqnarray}}
\def\bsplit{\begin{split}}
\def\esplit{\end{split}}

\def\nn{\nonumber}
\def\f{\frac}

\def\l{\left(}
\def\r{\right)}
\def\[{\left[}
\def\]{\right]}
\def\bs{\boldsymbol}
\def\ddx{\partial_x}
\def\ddxx{\partial^2_x}

\def\be{\begin{equation}}
\def\ee{\end{equation}}
\def\bea{\begin{eqnarray}}
\def\eea{\end{eqnarray}}

\def\nn{\nonumber}

\def\l{\left(}
\def\r{\right)}

\def\f{\frac}

\def\be{\begin{equation}}
\def\ee{\end{equation}}
\def\bea{\begin{eqnarray}}
\def\eea{\end{eqnarray}}
\def\bsplit{\begin{split}}
\def\esplit{\end{split}}

\def\nn{\nonumber}
\def\f{\frac}

\def\l{\left(}
\def\r{\right)}
\def\bs{\boldsymbol}
\def\ddx{\partial_x}

\begin{document}

\preprint{APS/123-QED}

\title{Excitability and travelling waves in renewable active  matter}

\author{M. Abhishek$^\star$, Ankit Dhanuka$^\star$, Deb Sankar Banerjee$^{\dagger,\ddagger}$ and Madan Rao$^\star$}
 \affiliation{$^\star$Simons Centre for the Study of Living Machines, National Centre for Biological Sciences-TIFR, Bangalore 560065\\
$^\dagger$Department of Chemistry, University of Chicago, Chicago, IL, 60637\\
$^\ddagger$The James Franck Institute, University of Chicago, Chicago, IL, 60637}




\date{\today}

\begin{abstract}
Activity and renewability are distinctive features of living matter, and constitute a new class of materials that we term {\it renewable active matter}. 
A striking example is the cell cytoskeleton, where myosin filaments bind to the actin meshwork,  apply contractile stresses and undergo continual stress/strain dependent turnover, thus acting as both force generators and sensors. As a consequence of nonreciprocity, arising from the independence of action and response, such living matter exhibits unusual mechanical  properties like,  segregation without attraction, 
fragility and force chains.
Here we show that the interplay between activity and turnover gives rise to mechanical excitability in the form of travelling waves and pulses, and spatiotemporal chaos. 
  We provide a systematic study of  the 
   nucleation, movement and shape of the travelling pulse, and present  a boundary layer analysis to establish the existence of 
{\it homoclinic orbits}. Our analytical results are supported by  detailed numerical analysis of the governing partial differential equations. This study has implications for the observed mechanical excitability in a variety of cellular contexts such as in isolated adherent cells and confluent cells within tissues.




\end{abstract}

\maketitle


\section{\label{sec:intro}Introduction}
\noindent

From ecosystems, organisms, to tissues and cells, the sustenance of living systems is contingent on the renewal of its constituents. In living cells, continual renewal is vital for the maintenance of intracellular patterning \cite{MARSHALL2020R544}. This is especially significant in the cell cytoskeleton, where turnover of actin filaments and myosin motors that bind to and apply contractile stresses on the actin meshwork, drive force patterning and sustained excitability.
Such cytoskeletal excitability is observed in a variety of cellular and developmental contexts~\cite{martin2009pulsed,martin2014apical,munjal2014actomyosin,munjal2015self,fernandez2011oscillatory,solon2009pulsed,meghana2011integrin,blanchard2010cytoskeletal}.


In this paper, we explore in detail, the origin and sustenance of the many modes of mechanical excitability, using an active hydrodynamic description of the renewable actomyosin cytoskeletal meshwork~\cite{Banerjee_2011,banerjee2011instabilities,banerjee2017actomyosin}. The actin cytoskeleton presents itself as a cell-spanning filamentous mesh held by crosslinkers,
that responds elastically to internal contractile stresses
generated by the ATP-driven force transduction of bound 
Myo-II minifilaments.
The Myo-II minifilaments continually unbind from the meshwork, often in a stress or strain dependent manner~\cite{kovacs2007load,fernandez2009myosin,mulla2022weak,hiraiwa2016role,weirich2021actin} with catch or slip bond response.
Thus Myo-II minifilaments are both {\it stress generators} and {\it stress sensors}, regulated by independent biochemical cycles. The ensuing 
mechanical nonreciprocity is the source of the spatially extended mechanical excitability reported here.

In a previous study~\cite{banerjee2017actomyosin}, we had formulated and analysed a set of hydrodynamic equations that described the dynamics  of apical actomyosin in germ-band cells, using a combination of linear stability and numerical analysis.
Here, we explore the nonlinear excitability of this system in detail, using a variety of analytical and numerical techniques. In particular, we conduct a detailed analysis of the activation and description of travelling pulses. The excitability observed in these situations are driven solely by  mechanical agents~\cite{munjal2015self}, as opposed to chemically-driven excitable systems, such as the Rho GTPase oscillator~\cite{michaux2018excitable,bement2015activator,staddon2022pulsatile}.

In Sect.\,\ref{sec:eqns}, we present a {\it nonreciprocal field theory}
 that describes the deformations of the active elastomer, subject to the binding-unbinding of the active stress generators. In Sect.\,\ref{sect:linear}, we give a brief recap of the linear stability analysis (details of which appear in~\cite{banerjee2017actomyosin}), with emphasis on the regime of excitability. A nonlinear analysis using a Galerkin approach is carried out in Sect.\,\ref{sect:nonlinear}, where we systematically arrive at various features of excitability, such as spontaneous oscillations, contractile instabilities and chaos. This leads us to a detailed study of the emergence and characteristics of travelling pulses in Sect.\,\ref{sect:spatioexcit},\, \ref{sect:nuctrans}, followed by 
an analysis of spatiotemporal chaos in Sect.\,\ref{sect:stchaos}. Such localised travelling pulse solutions appear in other
excitable systems such as the Fitzhugh-Nagumo model~\cite{fitzhugh1961impulses,nagumo1962active} and self-aggregation models such as the Keller-Segel model~\cite{keller1970initiation} to which our model bears close resemblance.


\section{Hydrodynamics of a renewable active elastomer}\label{sec:eqns}
The hydrodynamic equations for an active elastomer embedded in a fluid is written in terms of  -
(i) $\rho^a$, the mass density of the elastomer; (ii) $\bs u$, the displacement field of elastomer from its reference stress-free configuration; (iii) ${\rho}$, the density of active stress generators (e.g., 
bound myosin minifilaments); and (iv) ${\bs v}$, the solvent velocity. 

In general, both myosin and actin could undergo turnover, albeit at different time scales. Actin turnover can result in a solid to fluid transition as the  actin unbinding rate increases~\cite{banerjee2017actomyosin}. Here we will focus on myosin turnover alone, i.e.  we will be concerned with cellular contexts where the connectivity of the meshwork is strictly maintained, thus rendering it elastic at all times.

Ignoring the hydrodynamics of the embedding fluid, the overdamped dynamics 
of the elastomer is described by a set of partial differential equations (pde's) representing force, torque and myosin mass balance~\cite{Banerjee_2011,banerjee2011instabilities,banerjee2017actomyosin,solon2009pulsed},
\begin{equation}
\Gamma {\dot {\bs u}}  = \bs \nabla \cdot (\bs \sigma^e + \bs \sigma^a +\bs \sigma^{d})
\label{eq:mesh}
\end{equation}
\begin{equation}
\dot{\rho} + \bs \nabla \cdot (\rho \dot{\bs u})= D\bs\nabla^2\rho + {\cal S}_m\, ,
\label{eq:myosin}
\end{equation}
where
$\Gamma$ is the mesh friction, $D$ is the diffusion coefficient of bound myosin (acting as a regularizer for the pde \cite{eggers2015singularities}), and the turnover ${\cal S}_m$ describes the binding and unbinding of myosin filaments.
The total stress in Eq.\,\eqref{eq:mesh} is the 
sum of the elastic stress ($\boldsymbol \sigma^e$), viscous stress ($\boldsymbol\sigma^{d}$), and active stress (${\bs \sigma}^a$).

The elastic stress $\boldsymbol \sigma^e = \frac{\delta F}{\delta \boldsymbol \epsilon}$ is 
 derived from a free-energy functional $F[{\bs \epsilon}, \rho^a, \rho]
=\int d{\bf r} f_B$, where, $\boldsymbol\epsilon=(\nabla{\bs u}+(\nabla{\bs u})^T)/2$ is the linearized symmetric strain tensor, and
\begin{equation}
f_B = \frac{B}{2} \epsilon^2 + \mu |\widetilde{\boldsymbol{\epsilon}}|^2+ c\, \delta \rho^a \epsilon + \frac{1}{2} (\delta \rho^a)^2 \, .
\label{eq:elastic}
\end{equation}
Here $\epsilon$ is the compressive part of the strain and $\widetilde{\boldsymbol{\epsilon}}$ its deviatoric part, 
$B$ and $\mu$ are the bare compressive and shear moduli of the elastomer, dependent on cross-linker density. For simplicity we set the compressibility (coefficient of $(\delta \rho^a)^2$) to $1$, where 
$\delta \rho^a \equiv\rho^a -\rho_0^a$ is the deviation of the local mesh density from the uniform, unstrained equilibrium density $\rho_0^a$. We take the actin mesh density to be ``fast'' and
slaved to the local 
compressive strain, i.e., $\delta \rho^a =  - c \epsilon$.
The elastic stress is now given by,
\begin{eqnarray}
     \boldsymbol\sigma^{e} &=& 
 (B - c^2)\epsilon + 2\mu\widetilde{\boldsymbol{\epsilon}} \, .
\label{eq:elasticstress}
\end{eqnarray}
 
The dissipative stress is related to the strain rate of the elastomer, $\boldsymbol\sigma^{d}=\eta (\nabla{\dot{\bs u}}+(\nabla{\dot{\bs u}})^T)/2$, where $\eta$ is the mesh viscosity.
Since the mesh is isotropic at large scales~\footnote{There are  situations where we need to include anisotropic contributions to the active stress such as in the emergence of stress fibres~\cite{ayanstressfiber}}, the active stress $\bs \sigma^a$, only contributes to a pressure, which depends on $\rho$ and $\rho^a$ - it is reasonable to take it to increase and saturate with increasing myosin density $\rho$~\cite{marchetti2013hydrodynamics},
\be
{\bs \sigma}^a 
=  \f{\zeta_1 \rho}{1+\zeta_2 \rho}  \chi(\rho^a) \Delta \mu \,\bs I
\label{eq:activephen}
\ee
where $\Delta \mu$ is the fixed chemical potential difference during ATP hydrolysis which we take to be $\Delta \mu = 1$, $\chi(\rho^a)$ is a smooth, positive valued function of the mesh density, and $\bs I$ the identity matrix.
Parameters $\zeta_1>0$  and $\zeta_2 > 0$ ensure that the active stress is contractile~\cite{marchetti2013hydrodynamics}, which  draws in the surrounding material.
Taylor expanding $\chi(\rho^a)$ in Eq.\,(\ref{eq:activephen}) about $\rho_0^a$, 
$
\chi (\rho^a) = \chi_0(\rho_0^a) - c \chi_1(\rho_0^a) \epsilon +  c^2 \chi_2(\rho_0^a) \epsilon^2 + \ldots$, 
allows us to
recast
the active stress as
\be
\sigma^a =   \f{\zeta_1 \rho}{1+\zeta_2 \rho}  \l  \chi_0(\rho_0^a) - c \chi_1(\rho_0^a) \epsilon + c^2 \chi_2(\rho_0^a) \epsilon^2 
+ \ldots \r \, .
\label{eq:sigmaa}
\ee
Separating out the terms dependent on $\epsilon$ from those that depend 
 only on $\rho$, the
 effective elastic energy $\Phi(\boldsymbol{\epsilon})=\int_0^{\boldsymbol{\epsilon}}\boldsymbol{\sigma}(\boldsymbol{\epsilon}')\cdot d\boldsymbol{\epsilon}'$, can be expressed as 
\be
\Phi(\epsilon) = \frac{K_2}{2}\epsilon^2 + \frac{K_3}{3} \epsilon^3 + \frac{K_4}{4}\epsilon^4 + \mu |\widetilde{\boldsymbol{\epsilon}}|^2 \, ,
\label{eq:eff-free}
\ee
where the (non)linear coefficients depend on
myosin and actin density: 
$K_2 = B - c^2 -  \f{\zeta_1 \rho}{1 + \zeta_2 \rho} c \chi_1(\rho_0^a)$,
 $K_3 =  \f{\zeta_1  \rho}{1 + \zeta_2 \rho} c^2 \f{\chi_2(\rho_0^a)}{2}$ and $K_4 =  -\f{\zeta_1 \rho}{1 + \zeta_2 \rho} c^3 \f{\chi_3(\rho_0^a)}{6}$.
Since $\zeta_1>0$ and $\chi_1(\rho_0^a)>0$,  $K_2$ goes from being positive to negative as contractility increases. The signs of the other constants are:
$\zeta_2 > 0$, $\chi_2(\rho_0^a)>0$ and $\chi_3(\rho_0^a)<0$, so that $K_3$ and $K_4$ are always positive, the latter  ensuring that the  local compressive strain does not grow without bound, as a consequence of  steric hindrance,  filament rigidity or crosslinking myosin. 
As contractility increases, the effective elastic energy $\Phi(\epsilon)$ goes from having one minimum at $\epsilon=0$
with renormalised linear elastic moduli when $K_2>0$, to  multiple minima at $\epsilon\neq 0$ together with a nonlinear stabilization. This alteration of the effective free energy landscape is observed to be a generic feature of active  systems~\cite{maitra2020enhanced,sheshka2016rigidity,banerjee2017actomyosin,roychowdhury2024segregation,roychowdhury2024active,behera2023enhanced}.


Equation\,\eqref{eq:mesh} can be rewritten as,
\be
\label{eq:finalu}
\Gamma {\dot {\bs u}}   = \bs \nabla \cdot \l \Phi^{\prime}(\boldsymbol\epsilon) +\sigma^a(\rho)+ \eta \dot{\boldsymbol\epsilon} \r
\ee
where $\Phi^{\prime}\equiv \frac{\delta \Phi}{\delta \boldsymbol \epsilon}$ and 
$\sigma^a(\rho) = \f{\zeta_1  \Delta \mu \rho}{1+\zeta_2 \rho}  \chi_0(\rho_0^a)$.

Equation\,(\ref{eq:myosin})  for the bound myosin density
describes advection by the actin mesh velocity ${\dot {\bf u}}$ and 
turnover ${\cal S}_m=-k_u(\bs \epsilon) \rho + k_b \rho^a$ as a consequence of binding and unbinding. Most mechano-chemical proteins, and myosin minifilaments in particular, exhibit
a strain-induced unbinding which we take to be of the Bell-form,
$k_u(\bs \epsilon)= k_{u0} e^{{\bs \alpha}\cdot{\bs \epsilon}}$  
\cite{kovacs2007load}. 
The sign of $\alpha$ is  either positive or negative: 
$\alpha>0$ implies a local extension (compression) of the mesh will increase (decrease) the myosin unbinding rate, while $\alpha<0$ implies a local compression (extension) of the mesh will increase (decrease) the myosin unbinding rate. 
Keeping the response of myosin minifilaments in mind, we will refer to $\alpha >0$ as {\it catch bond} response, and $\alpha < 0$ as {\it slip bond} response.
The choice $\alpha=0$ implies that the myosin unbinding rate is a constant, independent of mesh deformation. The dynamics of bound myosin, after slaving the actin density to the compressive strain, is given by,
\be
\dot{\rho} + \bs \nabla \cdot (\dot{\textbf{u}} \rho) = D \bs\nabla^2 \rho - k_{u0} e^{{\bs \alpha}\cdot{\bs \epsilon}} \rho + k_b (1- c \epsilon) \,.
\label{eq:rhomyo}
\ee 
With  the hydrodynamic equations \eqref{eq:finalu} and \eqref{eq:rhomyo}, we impose boundary conditions on $u$, which can be either clamped (Dirichlet), stress-free (Neumann) or periodic, with $\rho$ obeying either no flux or periodic boundary conditions. We note that the distinct chemical networks governing the active force generation and force sensing (via myosin turnover), is at the root of the nonreciprocity of force and reaction exhibited by this system of equations.

Before going ahead, it is convenient to express the above equations in dimensionless form, where time and space are scaled as 
$tk_b \to t$ and $x/l \to x$, where $l = \sqrt{\frac{\eta}{\Gamma}}$. Using  the homogeneous myosin density $\rho_0$ to scale $\rho$, and $l$ to scale $u$, the rest of the parameters are redefined as, 
\begin{eqnarray}
\label{eq:nondimpar1}
    \frac{u}{l} \to u, \frac{\rho}{\rho_0} \rightarrow \rho , \frac{D}{k_b l^2} \rightarrow D \\
    \frac{k_{u0}}{k_b} \rightarrow k, \frac{B}{\Gamma k_b l^2} \rightarrow B, \frac{\zeta_1 \rho_0}{\Gamma k_b l^2} \rightarrow \zeta_1 \, ,
    \label{eq:nondimpar2}
\end{eqnarray}
leading to the following nondimensional equations, 
\begin{eqnarray}\label{eq:nondimfinal}
    \dot{\boldsymbol u} &=& \nabla \cdot (\Phi^{\prime}(\boldsymbol\epsilon) +\sigma^a(\rho)+ \dot{\boldsymbol\epsilon}) \\
    \label{eq:nondimrho}
    \dot{\rho} + \nabla \cdot (\dot{\boldsymbol u} \rho) &=& D \nabla^2 \rho + 1- c \epsilon - k e^{\alpha \epsilon}\, .
\end{eqnarray}

\section{Linear analysis of hydrodynamic equations\label{sect:linear}}
We extend the linear stability analysis carried out in~\citep{banerjee2017actomyosin} to highlight the region of excitability.
We linearize  Eqs.\,\eqref{eq:nondimfinal} and\,\eqref{eq:nondimrho} in deviations about the unstrained, homogeneous elastomer,
$\delta u = u - u_0$ and $\delta \rho = \rho - 1$, 
\begin{eqnarray}
\label{eq:linstabu}
 (1-\nabla^2) \delta \dot{u} & = & (B - c^2 - \frac{c \zeta_1 \chi_1(\rho_0^a)}{k}) \nabla^2 \delta u + \zeta_1\nabla \delta \rho \\
 \label{eq:linstabrho}
 \delta \dot{\rho} + \frac{1}{k} \nabla \delta \dot{u} & = & D \nabla^2 \delta \rho - (c+\alpha) \nabla \delta u - k \delta  \rho
\end{eqnarray}

Upon Fourier transformation,  ${f}(\textbf{q},t) = \frac{1}{(2\pi)^{d/2}} \int_{-\infty}^{\infty} f(\textbf{x},t) e^{-i \textbf{q}\cdot\textbf{x}}\,d\textbf{x}$, these linearized equations reduce to an eigenvalue equation,
\begin{equation}
\begin{split}
\begin{bmatrix}
\delta \dot{u} \\
\delta \dot{\rho}
\end{bmatrix}
=
\textbf{M} \begin{bmatrix}
\delta u \\
\delta \rho
\end{bmatrix}
\end{split}
\label{eq:lindyn}
\end{equation}
where the dynamical matrix 
\begin{equation} 
\textbf{M}=
\begin{bmatrix}
\frac{-q^2(B-c^2 - \frac{c \chi_1 \zeta_1}{k})}{1 + q^2} & \frac{-i q \zeta_1}{1 + q^2} \\
i q (c + \alpha) - \frac{i q^3(B - c^2 - \frac{\chi_1 c \zeta_1}{k})}{k(1 + q^2)} & - k - D q^2 + \frac{q^2 \zeta_1}{k (1 + q^2)}
\end{bmatrix}\, .
\end{equation}

The eigenvalues of this dynamical matrix can be written in the form $ \lambda_{\pm} = \frac{\lambda_a \pm \sqrt{\lambda_b}}{2 k^2 (1 + q^2)}$ where,
\begin{eqnarray}
\lambda_a & = & -k^3 + q^2 \biggl(-B k^2+c^2 k^2+c \zeta _1 k-D k^2-k^3+\zeta _1 k\biggr)-D k^2 q^4 + O(q^6) \nonumber \\
\lambda_b & = & k^6 + q^2 \biggl[2 k^4 \biggl(k \left(-B+c^2+D+k\right)+\zeta _1 (2 \alpha +3 c-1)\biggr)\biggr]\nonumber \\ &+&q^4 \biggl[k^2 \biggl( k^2 \left(2 k \left(-B+c^2+2 D\right)+\left(-B+c^2+D\right)^2+k^2\right)+2 \zeta _1 k \left(-B (c+1)+c \left(c^2+c+D+3 k\right)-D+2 \alpha  k-k\right) \nonumber \\&+&(c+1)^2 \zeta _1^2\biggr)\biggr] + O(q^6)
\end{eqnarray}

\begin {figure}
\centering
\includegraphics[width=11.5cm]{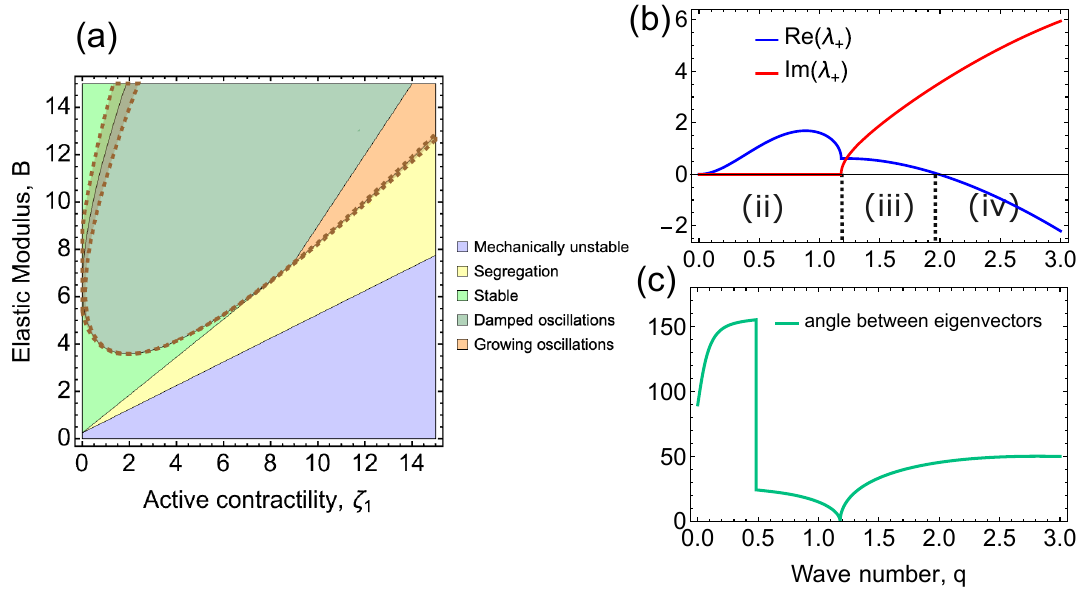}
\caption{Linear stability analysis. (a) Stability diagram for $q=2$ mode in  the elastic modulus ($B$) vs.\, active contractility ($\zeta_1$) plane.  The various phases are displayed in the legend. The shaded region bordered by the broken brown line indicates the region where the angle between eigenvectors is close to zero. Rest of the parameters are set to $c=0.1,D=1,k=1,\alpha =1$. (b) Typical dispersion curve at $B=7, \zeta_1 =12$ illustrates how the various phases, namely (ii) segregation, (iii) growing oscillations and  (iv) damped oscillations, are determined. (c) Plot of the angle between eigenvectors shows the existence of an exceptional point at $q\approx 1.2$. 
}
\label{fig:linPhaseDia}	  
\end{figure}

It is clear from above, that the instabilities are determined solely by $\lambda_+$, the maximum eigenvalue of $\mathbf{M}$. This allows us to classify the different phases in the stability diagram (Fig.\,\ref{fig:linPhaseDia}),
\begin{enumerate}[(i)]
\item Monotonically stable ($B'>0 , Re(\lambda_+)<0 , Im(\lambda_+)=0$),
\item Segregation and growth ($B'>0 , Re(\lambda_+)>0 , Im(\lambda_+)=0$),
\item Growing oscillations ($B'>0 , Re(\lambda_+)>0 , Im(\lambda_+)\neq0$),
\item Damped oscillations - stable ($B'>0 , Re(\lambda_+)<0 , Im(\lambda_+)\neq0$),
\item  Mechanically unstable ($B'<0$),
\end{enumerate}
where for convenience, we have 
set $B' := B - c^2 - \frac{c \zeta_1 \chi_1(\rho_0)}{k}$ as the activity renormalized linear elastic modulus. 

Since here we are interested in spatially extended excitability, we focus on the growing oscillatory phase. A study of the angle between the eigenvectors sheds light on the origin of oscillatory phases.  We first  note that $\mathbf{M}$ is
 non-Hermitian  due to nonreciprocity of the underlying equations; as a consequence, the eigenvectors along which the perturbations propagate are no longer orthogonal to each other and may even co-align for some parameter values (Fig.\,\ref{fig:linPhaseDia}(c)). 
These coaligned points are known as Exceptional Points~\cite{trefethen1993hydrodynamic} and have been shown to be a gateway to travelling states~\cite{fruchart2021non,you2020nonreciprocity,roychowdhury2024segregation}. In Fig.\,\ref{fig:linPhaseDia}(a), we demarcate the region where
the angle between the eigenvectors is close to zero, which is seen to adjoin the phase boundary separating oscillatory and non-oscillatory phases.
 

\section{Nonlinear analysis: Mode Truncation analysis}
\label{sect:nonlinear}
Here we ask whether nonlinearities in the hydrodynamic equations could stablilize the growing oscillatory modes, leading  to generic features of excitabililty such as pulsations, travelling waves and chaos. We note that
the nonlinear terms  arise from strain-dependent
unbinding of myosin, advection of bound myosin, and active stress. 
We systematically study the effects of these nonlinear terms using a Galerkin mode
truncation scheme~\cite{fletcher1984computational}.
\subsection{Galerkin Mode Truncation}
\label{subsect:galerkin}
In this method, we consider a `trial solution' of the partial differential equation (pde) as a small  perturbation about the spatially homogeneous solution. This  solution is expanded as a truncated series in a suitable basis (modes). Plugging this approximate trial solution into the pde, gives us a set of {\it nonlinear} ordinary differential equations (ode's) in the mode amplitudes.  

Consider a pde in the variable $\phi(x,t)$ 
\begin{equation}
{\dot \phi}(x,t) = f(\phi, \phi', \ldots, x,t)\, ,
\end{equation}
where dot and prime denote derivaties with respect to $t$ and $x$, respectively, and $f$ is in general, a  nonlinear function of its arguments.
Let $\phi_0$ be its homogeneous steady state solution. We rewrite the pde in terms of the perturbation $\delta \phi(x,t) = \phi(x,t) - \phi_0$,
\begin{equation}\label{eq:gengalerkin2}
\delta {\dot \phi}(x,t) = g(\delta \phi(x,t), \ldots, x,t) \, .
\end{equation}
We now assume a trial solution for $\delta \phi$  of the form, 
\begin{equation}
\delta \phi (x,t) = \sum_{n = 0}^{N} a_n(t) \psi_n(x) 
\end{equation} 
where $\psi_n(x)$ is a set of basis functions appropriate to the boundary conditions (here we will use a Fourier basis) and N is finite (mode truncation).  
Inserting  this in Eq.\,\eqref{eq:gengalerkin2} and projecting onto the basis functions, 
\begin{eqnarray}
\int_{0}^{L} \psi_n(x)\, \delta {\dot \phi}(x,t) \, dx  &=& \int_{0}^{L} \psi_n(x)\, g(\delta \phi(x,t), \ldots, x,t)\, dx \, ,
\end{eqnarray}
we obtain a set of nonlinear ode's in $a_n(t)$ (where $n=0,1, \ldots N$), which need to be solved for a given set of initial data~\cite{linhares2007galerkin}. This yields an approximate solution for $\phi(x,t)$ which improves with increasing $N$ \cite{linhares2007galerkin}.

We now apply the Galerkin mode truncation to the pde's\,\eqref{eq:nondimfinal} and\,\eqref{eq:nondimrho} in 1D, written in terms of the strain $\epsilon = \ddx u$ and $\rho$, and perturbed about the homogeneous, unstrained state, 
\be
\dot{\epsilon} = \partial_x^2\biggl((B-c^2) \epsilon + \zeta_1 \rho(1 - \zeta_2\rho) \chi(\rho^a)\biggr)
\label{eq:eps_1d}
\ee
\be
\dot{\rho} + \partial_x (\rho \partial_x\sigma) = D \partial_x^2 \rho  -  k(1 + \alpha \epsilon + \frac{\alpha^2}{2} \epsilon^2 + \ldots)  \rho - c \epsilon	    \,.
\label{eq:rho_1d}
\ee 

The nonlinear terms in the above equations correspond to advection, active stress $\sigma^a(\rho,\rho^a) = \zeta_1 \rho \, (1-\zeta_2\rho) \,\chi(\rho^a)$ (where $\chi(\rho^a)$ is given by 
Eq.\,\eqref{eq:sigmaa}) and unbinding, where we have expanded out the active stress and the Bell-form of unbinding, for analytical ease. 

We choose periodic boundary conditions in a region $x\in \[0, L\]$,
\begin{subequations}
\begin{align}
  [\epsilon]_{x=0} =  [\epsilon]_{x=L} \\
  [\rho]_{x=0} = [\rho]_{x=L}
\end{align}
\label{eq:1mode_bc}
\end{subequations} 
and expand $\epsilon(x,t)$ and $\rho(x,t)$ in a Fourier basis (which obeys Eq.\,\eqref{eq:1mode_bc}) truncated at N
\begin{equation}
 \epsilon(x,t) = \sum^{N}_{n=0} \epsilon_n(t) \cos(\frac{2n\pi x}{L})
\end{equation}
\begin{equation}
 \rho(x,t) = \sum^{N}_{n=0} \rho_{n}(t) \cos(\frac{2n\pi x}{L})
\end{equation}
Projection onto the basis functions result in a nonlinear coupled dynamical system in terms of the mode amplitudes ($\epsilon_n(t),\,\rho_{n}(t)$). 
\subsection{1-mode analysis}
\label{sect:onemode}
We begin with the first nontrivial mode ($N=1$) consistent with the chosen boundary condition, 
\bea
\epsilon(x,t) = \epsilon_1(t)\cos (\frac{2 \pi x}{L}) 		\nn\\
\rho(x,t) = \rho_1(t) \cos(\frac{2 \pi x}{L}) \,
\eea
Substituting this in Eqs.\,\eqref{eq:eps_1d} and\,\eqref{eq:rho_1d} and projecting onto $\cos(\frac{2 \pi x}{L})$, 
results in a dynamical system in the amplitudes $\epsilon_1(t)$ and $\rho_1(t)$,
\begin{figure}[h]
\centering
\includegraphics[width=14 cm]{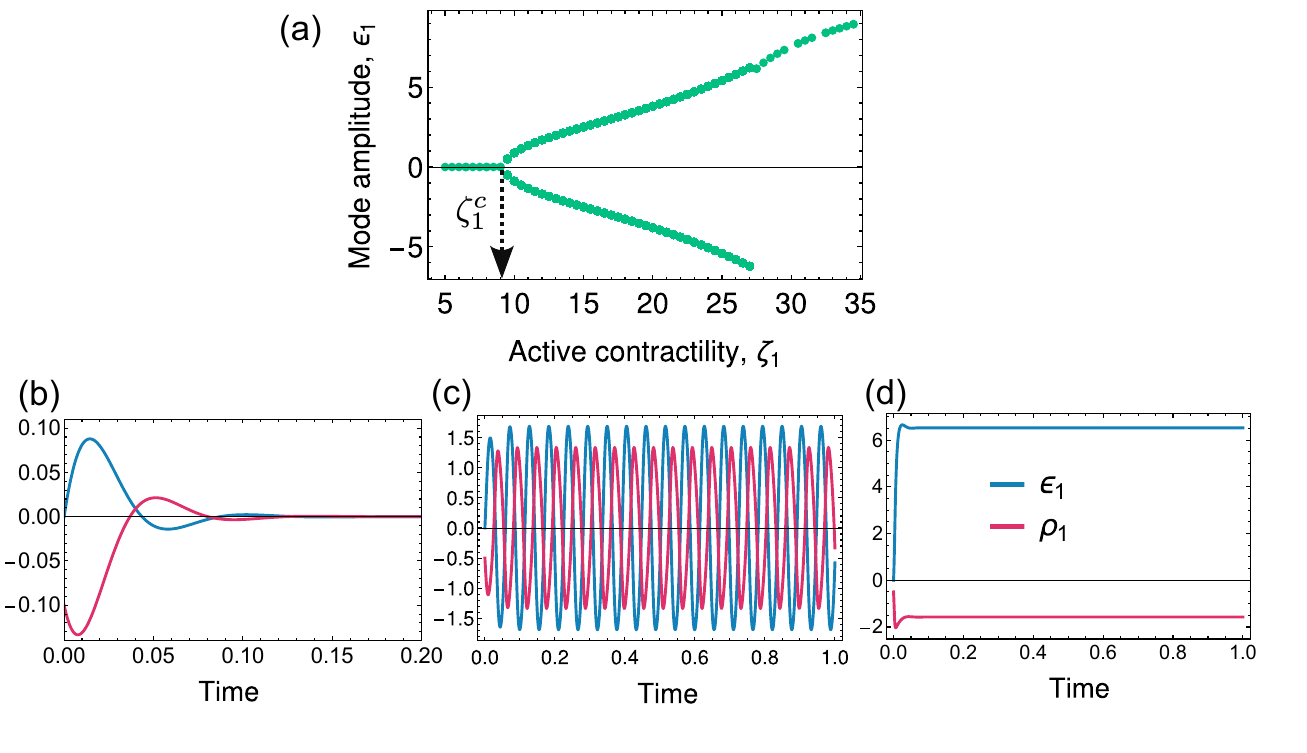} 
\caption[Stable, oscillatory and contracted steady state from one-mode analysis]{One mode analysis: (a) Bifurcation diagram of $\epsilon_{1}$ obtained by tuning the active contractility $\zeta_1$ with $\zeta_1^c = 9.3$ indicating the onset of oscillations. (b-d) Temporal dynamics of $\epsilon_1$ (blue line) and $\rho_1$ (red line) shows three distinct phases (b) Stable (when $\zeta_1 = 6$) and (c) Oscillatory (when $ \zeta_1 = 12$) and (d) Contracted (when $\zeta_1 =28$). Rest of the parameter values are set at $ B=5, c=0.1,\chi _0=1,\chi_1 = 1, \chi_2 =0.1,\chi_3=-0.01,D=1,k=1,\alpha =1,\zeta_2 =0.215$.}
\label{fig:1mode}
\end{figure}
\begin{eqnarray}
\dot{\epsilon_1} &=& \frac{\pi ^2}{k^2} \Bigg[-4 \biggl(k^2 \left(B-c^2\right)+c \zeta _1 \chi _1 \left(\zeta _2-k\right)\biggr)\epsilon _1+\frac{1}{2} c^3 \zeta _1 \chi _3 \left(k-\zeta _2\right)\epsilon _1^3- \zeta _1 k \left(k-2 \zeta _2\right) \left(4 \chi _0 \rho _1+ \frac{3}{2} c^2 \chi _2 \epsilon _1^2 \rho_1 \right)\nonumber \\&-&c \zeta _1 \zeta _2 k^2 \left(3 \chi _1 \epsilon _1 \rho_1^2+\frac{5}{12} c^2 \chi _3 \epsilon _1^3 \rho_1^2\right)\Bigg]
\label{eq:one_modeeps}
\end{eqnarray} 
\begin{eqnarray}
\dot{\rho_1} &=& -\biggl(\frac{k^2 \left(\alpha  (\alpha +1) k-4 \pi ^2 B\right)+4 \pi ^2 c^2 k^2+c \left(-4 \pi ^2 \zeta _1 \zeta _2 \chi _1+k^3+4 \pi ^2 \zeta _1 k \chi _1\right)}{k^3}\biggr)\epsilon_1 \nonumber \\  &-&\biggl(\frac{4 \pi ^2 d k^2+8 \pi ^2 \zeta _1 \zeta _2 \chi _0+k^3-4 \pi ^2 \zeta _1 k \chi _0}{k^2}\biggr) \rho_1 -\biggl(\frac{\pi ^2 c^3 \zeta _1 \chi _3 \left(k-\zeta _2\right)}{2 k^3}\biggr)\epsilon _1^3 + \biggl(\frac{\pi ^2 c^2 \zeta _1 \chi _2 \left(5 k-8 \zeta _2\right)}{2 k^2}\biggr) \epsilon _1^2 \rho_1 \nonumber \\ &+& \biggl(\frac{\pi ^2 c \zeta _1 \chi _1 \left(7 \zeta _2+2 k\right)}{k}\biggr) \epsilon _1 \rho_1^2 + \biggl(\frac{\pi ^2 c^3 \zeta _1 \chi _3 \left(13 \zeta _2-4 k\right)}{12 k}\biggr)\epsilon _1^3 \rho_1^2 -\pi ^2 \zeta _1 \zeta _2 \biggl(c^2 \chi _2 \epsilon _1^2+2 \chi _0\biggr) \rho_1^3
\label{eq:one_moderho}
\end{eqnarray}
At low values of active contractility ($\zeta_1$), the homogeneous, unstrained state ($\epsilon_1 = 0, \,\, \rho_1 = 0$) is a stable fixed point, Fig.\,\ref{fig:1mode}(b). At a critical value $\zeta_1^c$, a Hopf-bifurcation occurs and spontaneous bounded oscillations appear, Fig.\,\ref{fig:1mode}(c). When contractility is increased further, a new nontrivial fixed point at ($\epsilon_1 \neq0, \rho_1\neq0$) appears (Fig.\,\ref{fig:1mode}(d)) denoting a contracted state at high values of active contractility. These features are captured in a bifurcation diagram, Fig.\,\ref{fig:1mode}(a). 

In principle, the critical value $\zeta_1^c$ and the period of oscillations $T$ in its vicinity, may be calculated from an analysis of the eigenvalues $\lambda$ of the  linearized dynamical system, Eqs.\,\eqref{eq:one_modeeps} and \eqref{eq:one_moderho}, and setting 
$Re\{\lambda\}=0$.
For the parameters corresponding to Fig.\,\ref{fig:1mode}(a), we obtain
$\zeta_1^{c} = 9.27$ and  $T = 0.078$, which are in close agreement with the numerically estimated onset of bifurcation and period of oscillation (see Fig.\,\ref{fig:1mode}(a)).




Although the one-mode approximation already reveals spontaneous bounded oscillations, higher modes (smaller wavelength) are required to obtain more complex features of excitability, such as temporal chaos. 
\subsection{2-mode analysis}
\label{sect:twomode}
We now perform a two-mode ($N=2$) analysis  in $\rho$ and $\epsilon$,
\bea
\epsilon(x,t) &=& \epsilon_1(t) \cos(\frac{2 \pi x}{L}) + \epsilon_2(t) \cos(\frac{4 \pi x}{L}) 		\nn\\
\rho(x,t) &=& \rho_1(t) \cos(\frac{2 \pi x}{L}) + \rho_2(t) \cos(\frac{4 \pi x}{L})\, ,
\eea
consistent with the boundary conditions, which 
results in a 4-dimensional dynamical system in ($\epsilon_1,\epsilon_2,\rho_1,\,\rho_2$), that we display in Appendix \ref{app:galerkin}.
Our numerical analysis shows that, in addition to spontaneous oscillations, the two-mode bifurcation diagram in $\zeta_1-\epsilon_2$ space presents the following new features of excitability:\\
\begin{figure}[h]
\includegraphics[width=11cm]{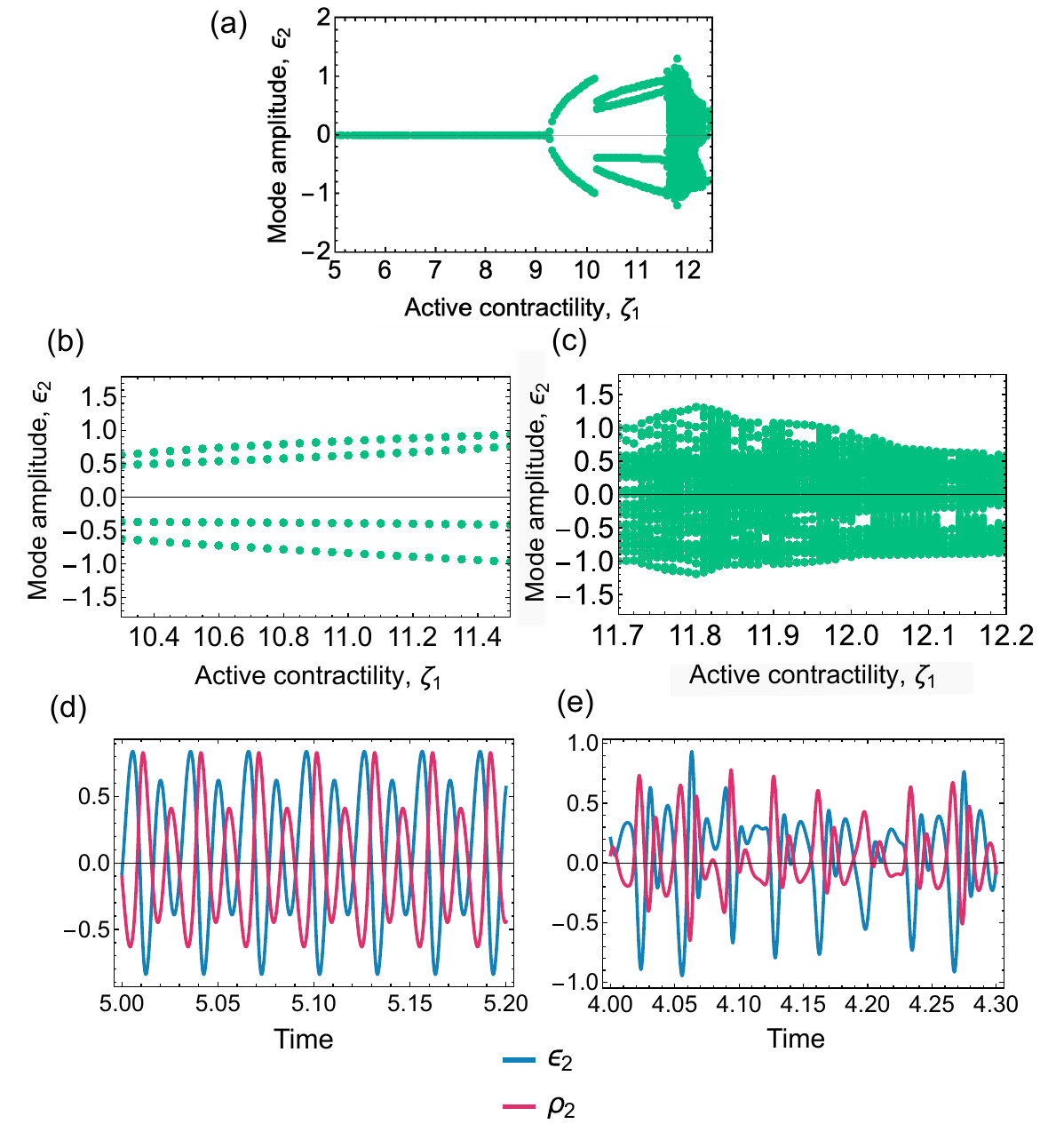} 
\caption{Two-mode analysis: (a) Bifurcation diagram of $\epsilon_{2}$ obtained by tuning the active contractility $\zeta_1$, showing additional phases such as period doubling. (b) Period doubling in $\epsilon_{2}$ when $\zeta_1$ is between $10.2$ and $11.5$, which transitions into (c) Space-filling at higher values of $\zeta_1$. (d-e) Time series of $\epsilon_2$ (blue line) and $\rho_2$ (red line) showing  (d) period-2 oscillations and (e) aperiodic oscillations leading to chaos. Rest of parameters set to $c=0.1,\chi_0=1,\chi_1=1,\chi_2=0.1,\chi_3 = -0.001, D=1,\zeta_2 = 0.215,B=5,k=1,\alpha =1$.}
\label{fig:2mode_phases}
\end{figure}\\
(i) {\it Period-doubling limit cycles:}
As before, for low values of $\zeta_1$, the homogeneous, unstrained state is a stable fixed point of the system   (Fig.\,\ref{fig:2mode_phases}(a)). As $\zeta_1$ increases, one sees spontaneous oscillations, followed by period-2 limit cycles (Fig.\,\ref{fig:2mode_phases}(b)\,and\,(d)) and period doubling. At higher values of  $\zeta_1$, this transitions into aperiodicity, raising the possibility of temporal chaos. \\
(ii) {\it Temporal Chaos:}
The chaotic phase becomes apparent as we zoom into the bifurcation diagram, Fig.\,\ref{fig:2mode_phases}(a). In the small region $\zeta_1 \in [11.7,12.2]$, the bifurcation diagram (Fig.\,\ref{fig:2mode_phases}(c)), shows space-filling behaviour, characteristic of temporal chaos. This is also confirmed by the aperiodic nature of the amplitude time series (Fig.\,\ref{fig:2mode_phases}(e)).

Our mode truncation analysis reveals that it is the driven and stabilizing nature of the nonlinear terms that is responsible for the emergence of temporal excitability in the form of  bounded period-doubling oscillations and temporal chaos. To investigate the spatiotemporal signatures of excitability, we perform a numerical and analytical study of the full nonlinear pdes.   
\section{Travelling states: Numerical treatment}
\label{sect:spatioexcit}
Numerical analysis of the full nonlinear spatially extended system 
reveals a rich phase diagram that include spatiotemporal excitable phases, such as 
travelling states.  In this section, we build the numerical phase diagram and make a detailed mathematical study of the travelling wave train solutions.

\label{subsec:numerical}

The dimensionless hydrodynamic equations are given by the pdes,
\be
\dot{\epsilon} = \partial_x^2 \sigma 
\label{eq:epsnum_1d}
\ee
and 
\be
\dot{\rho} + \partial_x (\rho \partial_x\sigma) = D \partial^2_x \rho - k e^{\alpha \epsilon} \rho + (1- c \epsilon) \,,
\label{eq:rhonum_1d}
\ee 
where the total stress $\sigma$ is written as a sum of elastic, active and dissipative stresses,
\begin{eqnarray}
\sigma  &=&  (B-c^2)\epsilon + \frac{\zeta_1\, \rho \chi(\rho^a)}{(1+\zeta_2\rho)} + \dot{\epsilon}\, ,
\label{eq:sigma}
\end{eqnarray}
and the form of $\chi(\rho^a)$ is obtained from Eq.\,\eqref{eq:sigmaa}.

We use a pseudospectral method in the Dedalus environment~\cite{burns2020dedalus} to numerically solve this system of pdes and obtain the full nonequilibrium phase diagram,  Fig.\,\ref{fig:numphasdiag}, in the $\zeta_1-B$ parameter space (see Appendix \ref{app:numerical} for details on the numerical method). 

Here, we display the phase diagrams for both catch bond ($\alpha>0$; Fig.\,\ref{fig:numphasdiag}(a)) and slip bond ($\alpha<0$; Fig.\,\ref{fig:numphasdiag}(b)) regimes. The catch bond regime shows the following  phases -- a segregated phase (where myosin patterns into high density and low density domains), a travelling pulse phase, an unusual coexistence phase of segregation and travelling waves, and a contractile collapse. 
This recapitulates the phases described in~\cite{banerjee2017actomyosin}. The slip bond regime, shows in addition, a wavetrain phase.
Typical travelling pulse and wavetrain solutions are shown in Movie \ref{mov:pulse}1 and Movie \ref{mov:pulse}3, respectively.  These travelling states are hallmarks of an excitable system. 


Each of these novel phases, arising due to nonreciprocity, deserve a separate detailed treatment.
The segregated phase is studied in detail in~\cite{roychowdhury2022emergence} while the unusual coexistence phase is being explored in an ongoing study. Here we will analytically examine the emergence and nature of the travelling phases. 
Together, these phases reveal the remarkable excitable properties of  renewable active matter, of which the actomyosin cytoskeleton is an exemplar.
\begin{figure}[h]
\center
\includegraphics[width=18.25 cm]{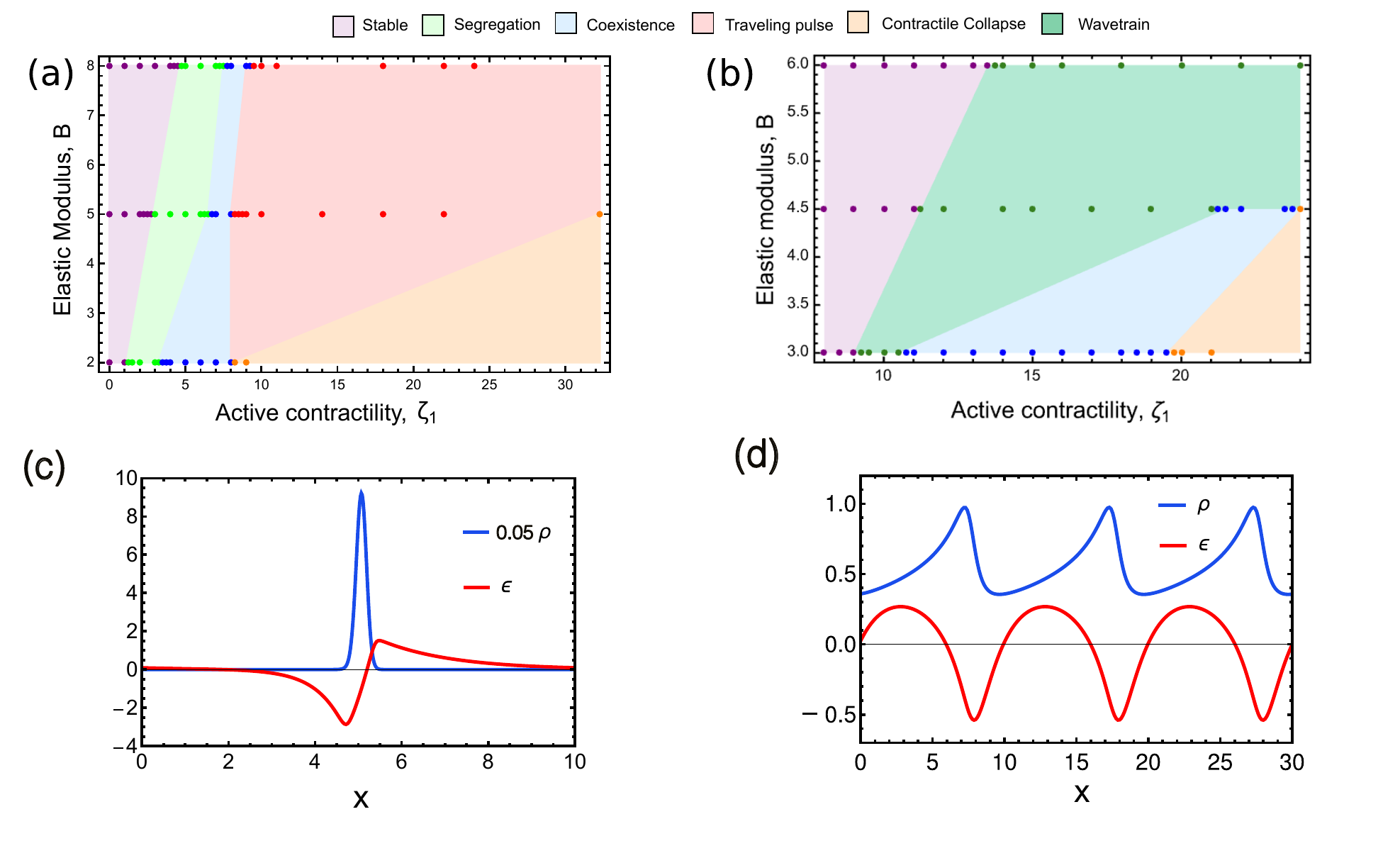} 
\caption{Numerical phase diagrams in the $\zeta_1-B$ plane for the (a) catch-bond regime, $\alpha=3$ and (b) slip-bond regime, $\alpha=-2.32$. Dots represent state-points where numerical data have been obtained, the colours identify the phases as depicted in the legend. (c,d) Snapshots of $\rho$ and $\epsilon$ in the right moving travelling pulse and travelling wavetrain, respectively. Note the asymmetry of the $\rho$ and $\epsilon$ profiles, which will be taken up in Sect.\ref{sect:nuctrans}. The other parameters are $ c=0.1, \chi_0=\chi_1=1, \chi _2=0.01, \chi_3=-0.001, D=1, \zeta_2 = 0.215, k=0.2$}
\label{fig:numphasdiag}
\end{figure}

\section{Travelling wavetrains: Analytical treatment}
\label{sec:nullwt}
The travelling states can be studied analytically, by noting that $\epsilon$ and $\rho$ will be functions of $\xi = x + v t$ alone, where $v$ is the velocity of the travelling wave.
The symmetry of Eqs.\,\eqref{eq:finalu} and \eqref{eq:rhomyo} allows for both right ($v<0$) and left ($v>0$) moving waves.
Transforming the pdes \eqref{eq:epsnum_1d} and\,\eqref{eq:rhonum_1d} to the comoving frame $\xi = x + v t$, gives the following 
pair of odes  
\begin{eqnarray} 
v\,\epsilon^{\prime} &=& \sigma^{\prime\prime}
\label{eq:excit1}
\end{eqnarray}
\begin{eqnarray}
\label{eq:excit2}
v \,\rho^{\prime} &=& -  (\rho \,\sigma^{\prime})^{\prime} + D \,\rho^{\prime\prime} + (1-c \epsilon) - k \, \,e^{\alpha \epsilon} \rho \, ,
\end{eqnarray}
where the derivative with respect to $\xi$ is denoted by a prime, and $\sigma$ is read from Eq.\,\eqref{eq:sigma}.

We will show from a phase space analysis that the above odes exhibit limit cycles here and homoclinic orbits (in Sect.\ref{sect:nuctrans}), as in other 
spatiotemporal excitable systems like the spatially extended Fitzhugh-Nagumo~\cite{winfree1980geometry} and slime-mold aggregation~\cite{article} models.

\begin{figure}[h]
\includegraphics[width=13.7 cm]{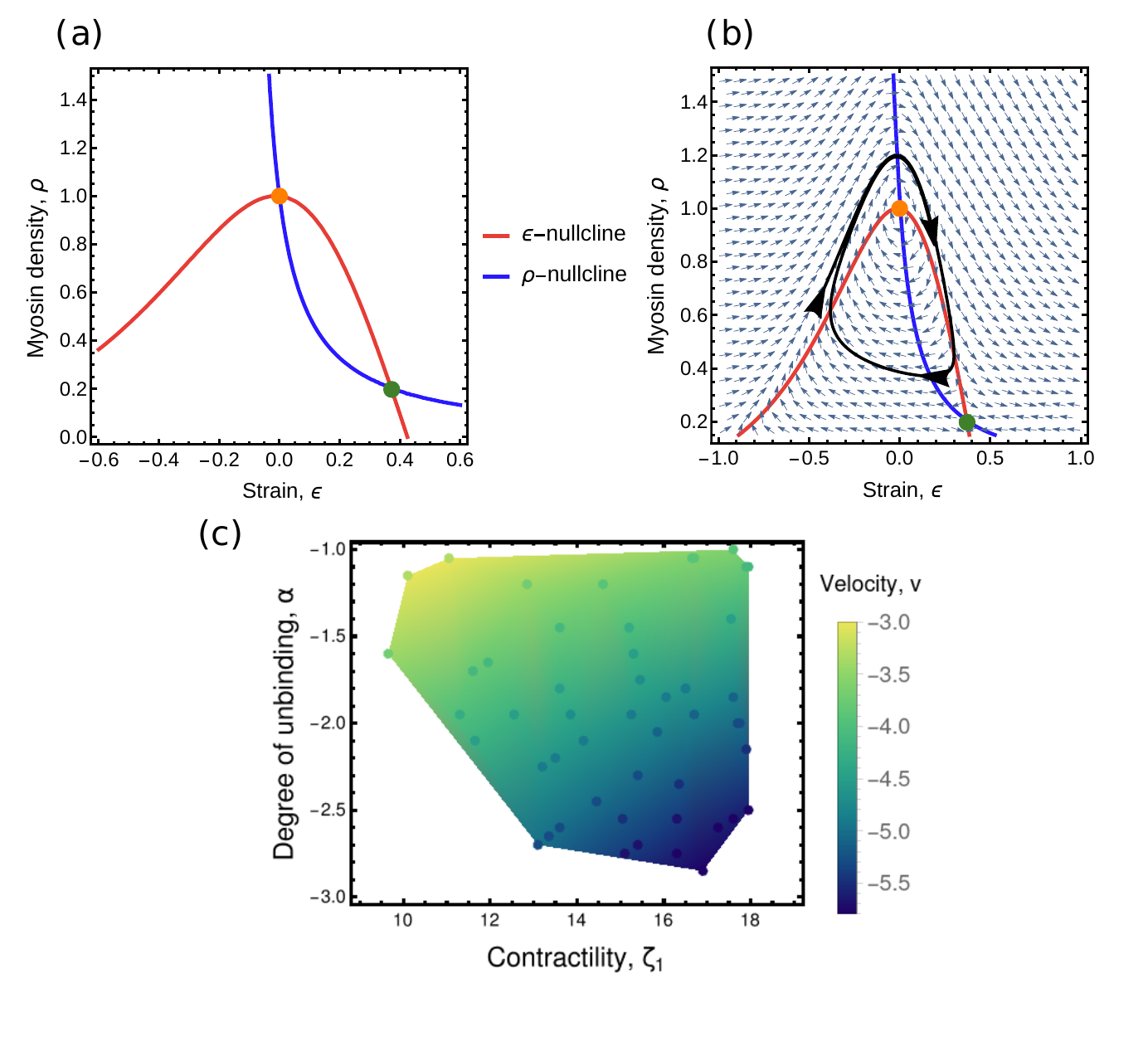} 
\caption[]{(a) Blue and Red curves are the nullclines of the system. Orange and Green dots indicate the fixed points. (b) Black curve indicates a limit cycle at $\alpha = -2.32$. Parameters values are for (a) and (b) are set to $ v = -4, B = 2.32, \zeta_1 = 9.68, \chi_0 = 1 ,k = 1, c = 1.$ (c) Phase diagram in $\zeta_1-\alpha$ space shows the  region where wavetrain solutions obtain, with the velocity indicated by the colour-bar. Dots indicate the state-points where the numerical simulations were performed. The boundary of the enclosed region is constructed by linear interpolation.}
\label{fig:wavetrainexcit}
\end{figure}

As seen in Movie\,\ref{app:numerical}3, the bound myosin with slip bond behaviour ($\alpha<0$) drives a wavetrain (see wavetrain phase in Fig.\,\ref{fig:numphasdiag}(b)).  Wavetrains are {\it spatial oscillatory} patterns moving with constant velocity. In the comoving frame, this would correspond to a limit cycle. To analyse the wavetrain, we start with Eqs.\,\eqref{eq:excit1},\,\eqref{eq:excit2}, ignore myosin 
 diffusion (which only acts as a regularizer), nonlinearity in active stress ($\chi^a$) and integrate Eq.\,\eqref{eq:excit1} once to obtain the first order odes, 
\begin{eqnarray}
(B-c^2)\epsilon' &=&  v  \epsilon  - (\frac{\zeta_1 \rho}{(1+\zeta_2\rho)})'
\end{eqnarray}
\begin{eqnarray}
v \rho' &=& - v (\rho \epsilon)' + (1- c \epsilon) - k \rho e^{\alpha \epsilon}
\end{eqnarray}
which may be rewritten as,
\begin{eqnarray}
\epsilon' &=&  -\frac{\frac{\zeta _1  \left(-c \epsilon +k \rho  e^{\alpha  \epsilon }+1\right)}{\left(\zeta _2 \rho +1\right){}^2}-v^2 \epsilon  (\epsilon +1)}{v (\epsilon +1) \left(B-c^2\right)-\frac{\zeta _1 \rho  v}{\left(\zeta _2 \rho +1\right){}^2}}
\label{eq:wtfineps}
\end{eqnarray}
\begin{eqnarray}
\rho' &=& \frac{\left(\zeta _2 \rho +1\right){}^2 \biggl(-k \rho  e^{\alpha  \epsilon } \left(B-c^2\right)+B c \epsilon -B-c^3\epsilon+c^2+\rho  v^2 \epsilon \biggr)}{\zeta _1 \rho  v -v (\epsilon +1) \left(B-c^2\right) \left(\zeta _2 \rho +1\right){}^2} \, .
\label{eq:wtfinrho}
\end{eqnarray}
The nullclines of this dynamical system intersect at two fixed points (orange and green in Fig.\,\ref{fig:wavetrainexcit}(a)). Inserting the appropriate  parameter values  where wavetrain solutions are obtained numerically (Fig.\,\ref{fig:numphasdiag}\,(b)) into Eqs.\,\eqref{eq:wtfineps}\,and\,\eqref{eq:wtfinrho}, we plot the phase portrait around the fixed points (Fig.\,\ref{fig:wavetrainexcit}(b)) which indicates a limit cycle. Fig.\,\ref{fig:wavetrainexcit}(c) shows the region in $\zeta_1-\alpha$ plane corresponding to wavetrain solutions, whose velocity $v$ is indicated by the colour-bar.     


While our analysis of wavetrains is for $\alpha<0$, such wavetrain solutions are also obtained for $\alpha<\alpha_{max}$ where $\alpha_{max}$ is a small positive value, as shown in~\cite{banerjee2017actomyosin}. 
However we note that myosin filaments that show catch bond behaviour ($\alpha>0$) at small strains must eventually slip at larger strains. Such situations will also show wavetrain solutions.

\section{Travelling pulses: Analytical Treatment\label{sect:nuctrans}}
In Fig.\,\ref{fig:numphasdiag}(a) and Movie \ref{app:numerical}1, we saw that the catch bond ($\alpha>0$) regime of Eqs.\,\eqref{eq:epsnum_1d} and \eqref{eq:rhonum_1d} shows a travelling pulse phase. Here we would like to study the onset of this travelling pulse as contractility ($\zeta_1$) is increased.
We numerically solve the pdes \eqref{eq:epsnum_1d} and\,\eqref{eq:rhonum_1d} using initial conditions where the myosin density is uniform with superimposed small fluctuations. 
We find that after some time, small undulations of myosin density nucleate, grow  and eventually move as a stable travelling pulse both in $\rho$ and $\epsilon$. Motivated by this, we pose two key questions:
\begin{itemize}
    \item At what value of contractility ($\zeta_1$) does the homogeneous steady state myosin profile become unstable to small perturbations? 
    This would correspond to a nucleation instability. 
    \item At what value of contractility ($\zeta_1$) does a growing myosin profile settle into a stable pulse travelling with constant velocity? This would correspond to the onset of translational instability. 
    \end{itemize}
To address these, we consider a simplified version of the equation for $\epsilon$ by setting $\chi^a = 1$, that reproduces the travelling pulse phase seen in Sect.\,\ref{subsec:numerical},
\begin{eqnarray}
\label{Eq:nucleation1}
 {\dot {\epsilon}}  &=& \ddx^2 \sigma
\label{eq:simplifiedeps}
\end{eqnarray}
where, 
\begin{eqnarray}
\sigma  &=&  (B-c^2)\epsilon + \frac{\zeta_1 \rho}{(1+\zeta_2\rho)} +  \dot{\epsilon}
\label{eq:simplifiedrho}
\end{eqnarray}
The dynamics for bound myosin is given by, 
\begin{eqnarray}
\label{Eq:nucleation4}
\dot{\rho} &=& -\partial_x (\rho \partial_x\sigma) + D \ddxx \rho +  (1- c \epsilon) - k  e^{\alpha \epsilon} \rho
\end{eqnarray}
We expand the dynamical variables in Hermite-Gaussian (H-G) polynomials \cite{abramowitz1948handbook}. This choice is supported by two assertions. Firstly, any arbitrary function can be expanded as a linear combination of H-G functions because they form a complete set of orthonormal basis functions. Secondly, numerically obtained profiles of both $\rho$ and $\epsilon$ in the travelling pulse (Fig.\,\ref{fig:numphasdiag}(c)) show that they can be approximated by a few Hermite-Gaussian functions such as, $e^{-x^2} , x e^{-x^2},\ldots$.
Projecting the dynamical variables onto these basis functions and integrating, we obtain a set of odes in the mode amplitudes. 
For analytical tractability, it is convenient to expand the active stress $\frac{\zeta_1\rho}{1+\zeta_2 \rho} \approx \zeta_1 \rho (1-\zeta_2 \rho + ...)$
and retain only the lowest order nonlinearity. 

\subsection{Nucleation instability}
\label{nucinstab:sec}
Explicitly, we expand $\rho$ and $\epsilon$ as,  
\begin{eqnarray}
\rho(x,t) &=& r+ \sum_{n = 0}^{\infty} \rho_{n}(t) \psi_n(x) 
\end{eqnarray}
\begin{eqnarray}
\epsilon(x,t) &=& \sum_{n = 0}^{\infty} \epsilon_{n}(t) \psi_n(x) 
\end{eqnarray} 
where $r = k_b/k_{u0} = 1/k$ is the steady state myosin density, $\epsilon_n(t)$ and $\rho_n(t)$ are the mode amplitudes and the H-G basis functions are given by,  
\begin{eqnarray}
    \psi_n(x)  = (2^n n! \sqrt{\pi})^{-\frac{1}{2}} e^{\frac{x^2}{2}} \frac{d^n}{dx^n} e^{-x^2} \,.
\end{eqnarray}
To inspect the onset of nucleation, we restrict the expansion to zeroth mode ($n=0$), 
\begin{eqnarray}
\epsilon(x,t) &=& \epsilon_{0}(t) \psi_0(x)
\label{eq:nucassumpeps}
\end{eqnarray}
\begin{eqnarray}
\rho(x,t) &=&  r + \rho_{0}(t) \psi_0(x)
\label{eq:nucassumprho}
\end{eqnarray} 
Inserting this into Eqs.\,\eqref{Eq:nucleation1} and \eqref{Eq:nucleation4} and projecting in the $\psi_0$-direction (i.e., multiplying by $\psi_0$ and integrating over $x$ from $-\infty$ to $+\infty$) leads to the amplitude equations,
\begin{figure}[htp]
\centering
\includegraphics[width=13.5 cm]{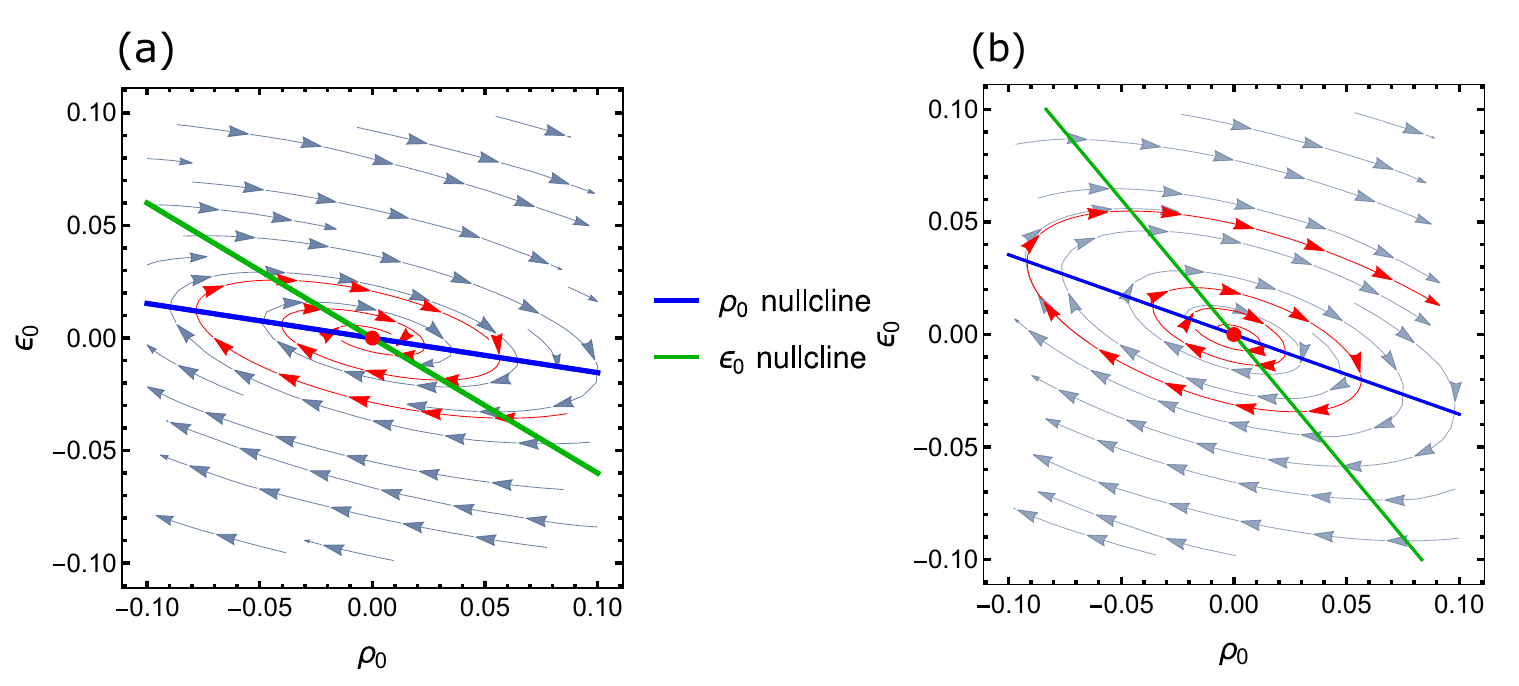} 
\caption[]{Nucleation instability as contractility $\zeta_1$ is increased. Phase plane dynamics in ($\rho_0, \epsilon_0$) shows intersection of the nullclines (blue/green) resulting in (a) Stable fixed point at (0,0) for low contractility $\zeta_1 = 5$, and (b) Unstable fixed point at (0,0) for high contractility $\zeta_1=10$  beyond the threshold $\zeta_{1}^{nuc} =6.38$ via a bifurcation. Other parameters are set at $k = 1$, $D = 0.5$, $\zeta_2=0.2$, $B=5$, $\alpha=0.2$ and $r=1$.}
\label{fig:nucleation}
\end{figure}   
\begin{eqnarray}
    \dot{\epsilon_0} &=& \frac{-(B-c^2) \epsilon_0 - \zeta_1 \rho_0 + 2 r \zeta_1 \zeta_2 \rho_0}{3}
    \label{eq:epsnucleation}
\end{eqnarray}
\begin{eqnarray}
    \dot{\rho_0} &=& -(\alpha \epsilon_0 + \rho_0) + \frac{\frac{6 B \pi \epsilon_0}{k} -2 D \pi \rho_0 + \frac{2 \pi \zeta_1 \rho_0}{k} - \frac{4 \pi \zeta_1 \zeta_2 \rho_0}{k^2}}{4 \pi}
    \label{eq:rhonucleation}
\end{eqnarray}

At low values of contractility ($\zeta_1$), the fixed point of this dynamical system is a stable spiral node which implies that perturbations in $\epsilon_0$ and $\rho_0$ will flow into fixed point Fig.\,\ref{fig:nucleation}(a). Above a threshold $\zeta_1^{nuc}$, the fixed point undergoes a crossover into an unstable spiral node as shown in Fig.\,\ref{fig:nucleation}(b).
We find this threshold $\zeta_1^{nuc}$ using an eigenvalue analysis of the linear dynamical system, Eqs.\,\eqref{eq:epsnucleation} and \eqref{eq:rhonucleation}. The eigenvalues are of the form, $\lambda=\lambda_1 \pm i\lambda_2$. The onset of nucleation instability occurs when $\lambda_1$ changes sign from -ve to +ve, i.e. when $\lambda_1 = 0$. 
Using this, we obtain the critical value,
\begin{eqnarray}
    \zeta_1^{nuc} &=& \frac{k^2 (2 B+3 D)}{-6 \zeta _2+72 k^3+k} \,,
\end{eqnarray}
above which small undulations of $\epsilon$ and $\rho$ will nucleate and begin to grow. 


In this parametrization (Eqs.\,\eqref{eq:nucassumpeps} and \eqref{eq:nucassumprho}) of $\epsilon$ and $\rho$, we do not find any nontrivial intersection point of the nullclines. This would suggest  unbridled growth of $\epsilon$ and $\rho$ beyond the nucleation threshold ($\zeta_1^{nuc}$), whereas, numerical analysis informs us that nucleated domains transform into a travelling pulse with constant shape and velocity. This can be captured by extending the basis functions, as shown in the following subsection.  
\subsection{Translation instability}

A travelling pulse can be analytically described as a shape-invariant profile moving with a constant velocity. For this, we extend the H-G basis as $\psi_n (x) \to \psi_n[x + a(t)]$, where $a(t)$ is a time dependent phase, whose time derivative will be identified with the travelling speed $v$. It is clear from the numerically obtained snapshots (Fig.\,\ref{fig:numphasdiag}(c)), that the profiles $\rho$ and $\epsilon$ are asymmetric, implying contributions from higher order modes. Expanding $\epsilon$ upto $n=2$ and $\rho$ upto $n=1$,
\begin{eqnarray}
    \epsilon (x,t) &=& \epsilon_0(t)\psi_0[x+a(t)] + \epsilon_1(t)\psi_1[x+a(t)] +  \epsilon_2(t)\psi_2[x+a(t)]
\end{eqnarray}
\begin{eqnarray}
    \rho (x,t) &=& r + \rho_0(t)\psi_0[x+a(t)]+ \rho_1(t)\psi_1[x+a(t)] \,,
\end{eqnarray}
and projecting the pde system onto ($\psi_0,\psi_1,\psi_2$), results in a 6-dimensional coupled dynamical system in ($\epsilon_0(t),\epsilon_1(t),\epsilon_2(t),\\\rho_0(t),\rho_1(t),a(t)$). This high-dimensional dynamical system  may be reduced further by appealing to the following considerations:

(i) Since in Eq.\,\eqref{Eq:nucleation1} the dynamics of $\epsilon$ is conserved, there exists an algebraic relation between its even mode amplitudes $\epsilon_0(t)$ and $\epsilon_2(t)$,
\begin{eqnarray}
    \pi^{\frac{1}{4}} [\epsilon_0(t)+\sqrt{2} \epsilon_2(t)] = s
\end{eqnarray}
where $s$ is the initial value of the total strain. This allows us to eliminate $\epsilon_2(t)$.

(ii) Existence of a travelling pulse corresponds to finding a nontrivial stable fixed point (which implies an invariant shape) with the constraint  $\dot{a}(t) = v \neq 0$. 

(iii) At the fixed point, $\dot{\epsilon_0}=\dot{\epsilon_1}=0$, which provides an algebraic relation for $\epsilon_0$ and $\epsilon_1$ in terms of $\rho_0$ and $\rho_1$, which can be used in the dynamical equations for $\rho_0$ and $\rho_1$. 

Together these reduce to a 2 dimensional dynamical system, 
\begin{eqnarray}
\label{eq:translation1}
    \dot{\rho_0} = g_0(\rho_0,\rho_1)\\
    \dot{\rho_1} = g_1(\rho_0,\rho_1)
    \label{eq:translation2}
\end{eqnarray}
Intersections of nullclines of this dynamical system determine its fixed points. The flows around the fixed points are given by $g_0(\rho_0,\rho_1)$ and $g_1(\rho_0,\rho_1)$.  

\begin{figure}[h]
\centering
\includegraphics[width=13.5 cm]{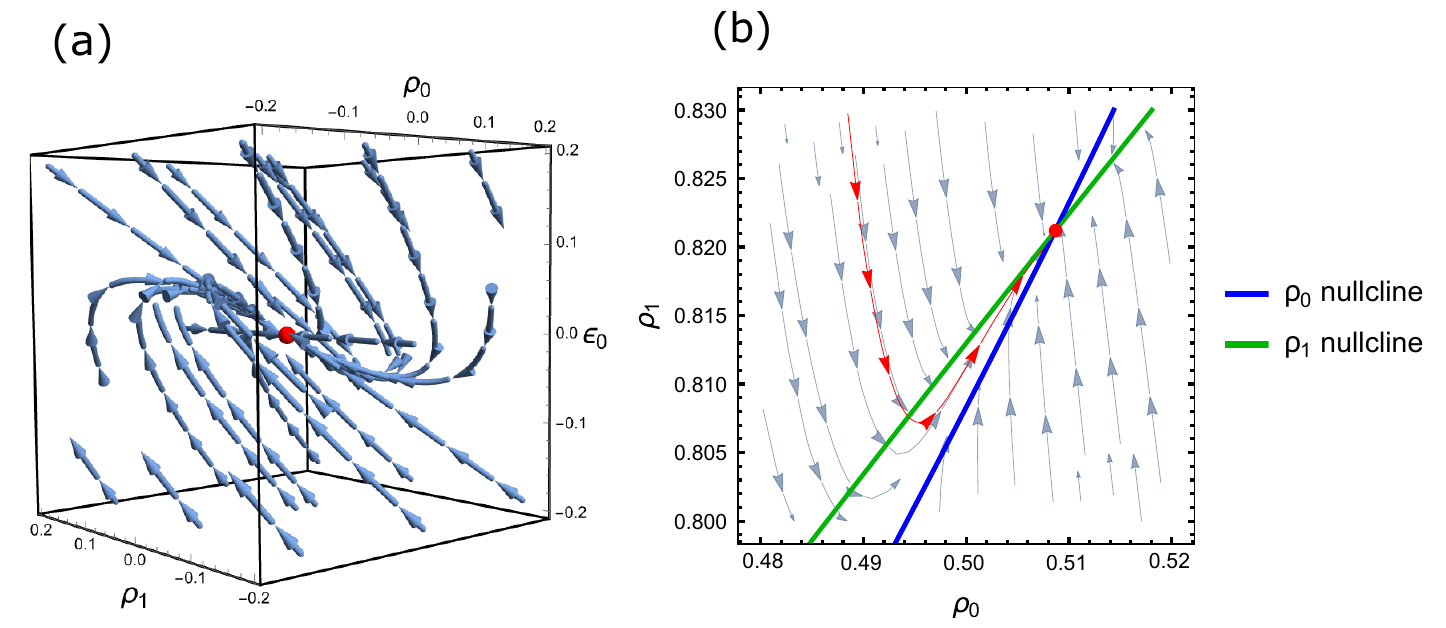} 
\caption[]{Translation instability as contractility $\zeta_1$ is increased. (a) At low $\zeta_1$($=5$), phase space analysis of the reduced dynamical system in $\rho_0, \rho_1, \epsilon_0$ shows the flows (blue trajectrories) into the trivial stable fixed point $(0, 0, 0)$ (red dot). This gets unstable beyond a critical threshold as noted in Fig.\,\ref{fig:nucleation}(b). (b) At contractility $\zeta_1 = 20$ beyond the threshold $\zeta_1^{trans}=16$, the  nullclines of Eqs.\,\eqref{eq:translation1},\,\eqref{eq:translation2} have a nontrivial intersection 
at $(\rho_0^*=0.822, \rho_1^*=0.509)$,
giving rise to flows (red trajectory) towards a stable fixed point (red dot) associated with a finite travelling speed $\dot{a}=1.6$.
 Other parameters are set at $k=1$, $D = 0.5$, $\zeta_2=0.2$, $B=5$, $\alpha=0.2$ and $r=1$.}
\label{fig:transphaseplane}
\end{figure}

\begin{figure}[h]
\centering
\includegraphics[width=12cm]{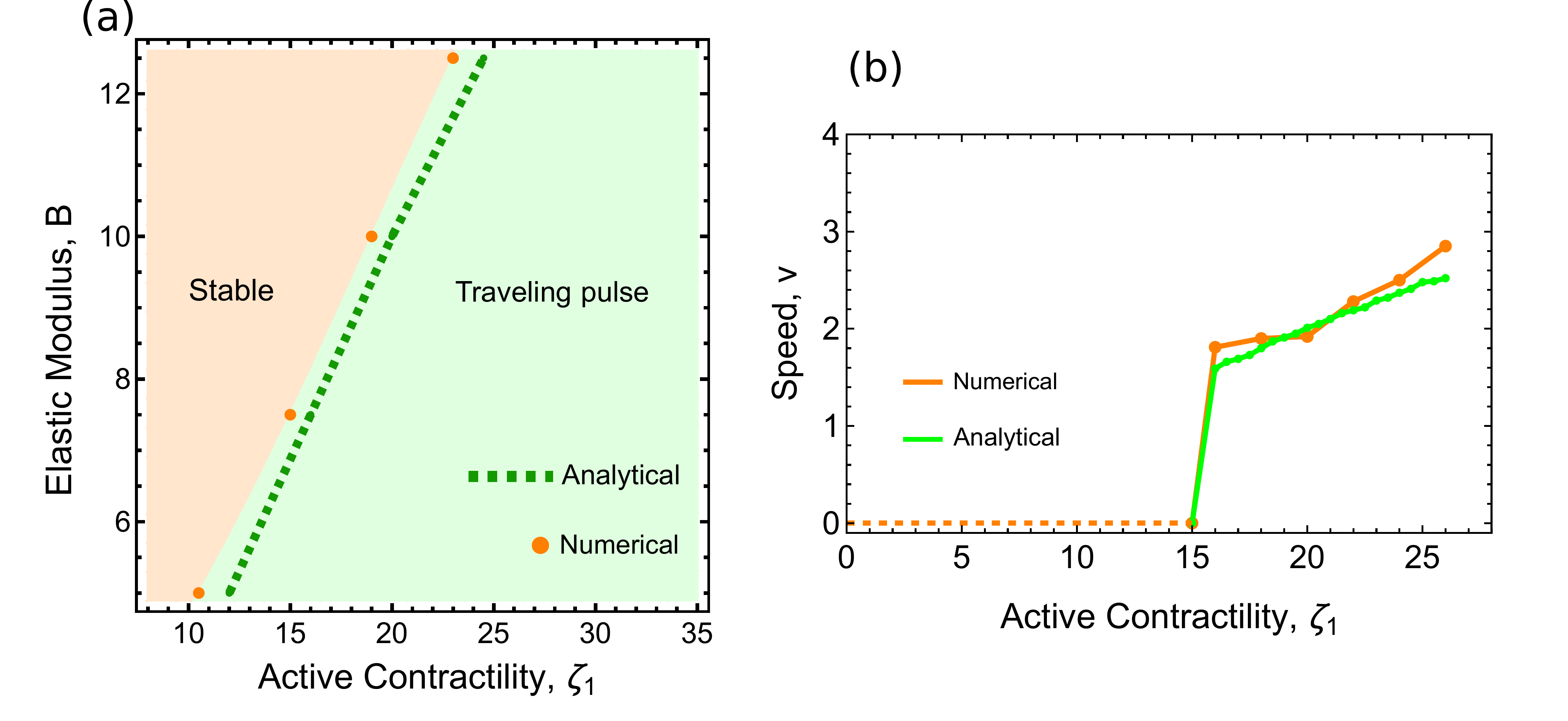} 
\caption[]{Dynamical phase transition. (a) Dynamical phase diagram in the plane of contractility ($\zeta_1^{trans}$) and elastic modulus ($B$) showing the onset of the travelling pulse obtained analytically (broken green line) and numerically (orange dots). (b) Plot of the travelling pulse velocity versus contractility ($\zeta_1$) shows a discontinuous dynamical phase transition, obtained analytically 
(green curve) and numerically (orange dots), for
$B=7.5$. Rest of the parameter values are set at $k=1$, $D = 0.5$ , $\zeta_2=0.2$, $\alpha=0.2$ and $r=1$.}
\label{fig:thresholdvelfig}
\end{figure}

At low values of contractility $\zeta_1$, the nullclines only intersect at $(0,0)$ corresponding to a stable fixed point (Fig.\,\ref{fig:transphaseplane}(a)), and the system is in the homogeneous stable phase (Fig.\,\ref{fig:thresholdvelfig}(a)). Above a threshold contractility ($\zeta_1^{trans}$), the nullclines have an intersection at $(\rho_0^*,\rho_1^*) \neq (0,0)$ and the flow around this fixed point is stable (red curve in Fig.\,\ref{fig:transphaseplane}(b)). This is associated with a finite velocity $\dot{a}$ and corresponds to the travelling pulse phase (Fig.\,\ref{fig:thresholdvelfig}(a)). Using the fixed point value, we may derive an analytic form for the velocity $\dot{a}$ of the travelling pulse.
This shows a discontinuous dynamical phase transition with the speed $v \sim (\zeta_1 - \zeta_1^{trans})$, beyond the threshold contractility (Fig.\,\ref{fig:thresholdvelfig}(b)). On monitoring the nature of the intersection of the nullclines in a high dimensional phase space, as the contractility $\zeta_1$ is varied, we find no sign of hysteresis across the phase boundary.


We compare this theoretical prediction with experimental studies of pulsation and flows in the apical actomyosin cortex of germband cells in the {\it Drosophila} embryo. To convert the dimensionless parameters used in the theory to physical units, we obtain the unit
of length $l = \sqrt{\eta/\Gamma}$ from the measured actin mesh size $0.5 \mu\text{m}$ \cite{rauzi2010planar}, and  the unit of time  $t = k_b^{-1}$, from  myosin FRAP measurements $k_b = 0.2 \text{s}^{-1}$ \cite{munjal2015self}. Using this in Eqs.\,\eqref{eq:nondimpar1} and \eqref{eq:nondimpar2} allows us to convert the nondimensional bulk modulus $B = 5$ and myosin contractility $\zeta_1 = 5$ to physical units $B=\zeta_1 \approx 42 \,\text{Pa}$, which are consistent with experimental values obtained from a study of the mechanical properties of cells in early drosophila embryo \cite{bambardekar2015direct}.


For the above set of parameters, our above analysis gives a pulse speed $v \approx 1$. This compares well with the value $v = 0.08 \mu\text{m/s}$ reported in \cite{rauzi2010planar,banerjee2017actomyosin}.


The appearance of a travelling pulse using a phase plane analysis in the truncated H-G basis, provides the motivation for establishing the existence of a homoclinic orbit in the comoving frame of the dynamical equations, which we take up in the next section.

\subsection{Travelling pulse appears as a homoclinic orbit via a boundary layer}
So far, we have investigated the conditions for the {\it onset} of a travelling pulse, using an approximate analysis based on an expansion in the truncated H-G basis. We now focus on its characteristics, such as its detailed profile and dynamics. What is our physical picture of the travelling pulse in strain and myosin density, especially since the equations describe the dynamics of myosin that is bound to the elastic mesh? There are two ways of thinking about this --
\begin{itemize}
    \item One is to start with a uniform density of bound myosin, set by the binding/unbinding rates. A local enhancement in bound myosin density will lead to a local compression, which can travel as a joint pulse of the compressive strain and myosin, akin to the propagation of a sound pulse. This can happen even in the absence of myosin binding and unbinding.
    \item The second is to allow (stochastic) turnover, with the myosin flux associated with binding and unbinding {\it being equal in the mean}. Thus averaged over the binding-unbinding cycle, the net myosin density is a constant. This homeostatic feedback is consistent with experiments in the {\it Drosophila} germ-band tissue  that show a ROCK pulse accompanying the travelling myosin pulse \cite{munjal2015self,banerjee2017actomyosin}.
\end{itemize}
Both these possibilities are mathematically consistent with the fact that for the travelling pulse to propagate as an invariant shape with constant net myosin density, the dynamics of bound myosin density must be conservative, at least in the mean.
We favour the second possibility as being biologically more relevant, since it is robust to fluctuations arising from myosin turnover.

With this interpretation, we can omit the non-conservative terms in the myosin density equation in our analysis of travelling pulse. 
This is indeed supported by our numerics. 
We now look for homoclinic orbits by transforming the pdes \eqref{Eq:nucleation1} and\,\eqref{Eq:nucleation4} to
the comoving frame $\xi = x+vt$ (primes are derivatives with respect to $\xi$),

\begin{eqnarray}
  v\,\epsilon^{\prime} &=& \biggl((B-c^2)\,\epsilon + \frac{\zeta_1 \rho}{(1+\zeta_2\rho)} + v\epsilon'\biggr)^{\prime\prime}
\label{eq:tpnullclineeps}
\end{eqnarray}
\begin{eqnarray}
v\,\rho^{\prime} &=& - v \,(\rho \,\epsilon)^{\prime} + D \, \rho^{\prime\prime}
\label{eq:tpnullclinerho}
\end{eqnarray} 

By doing this, $v$, which can be positive or negative, appears {\it parametrically} in the dynamical equations. Note that the above equations are invariant under the joint transformation $\xi \to - \xi$ and 
$v \to -v$; thus for every solution moving with a fixed $v$ there exists a parity-transformed solution moving with $-v$.

\subsubsection{Naive phase-plane analysis} 
\label{sect:tpulsenc}

\begin{figure}[h]
\center
\includegraphics[width=12.5 cm]{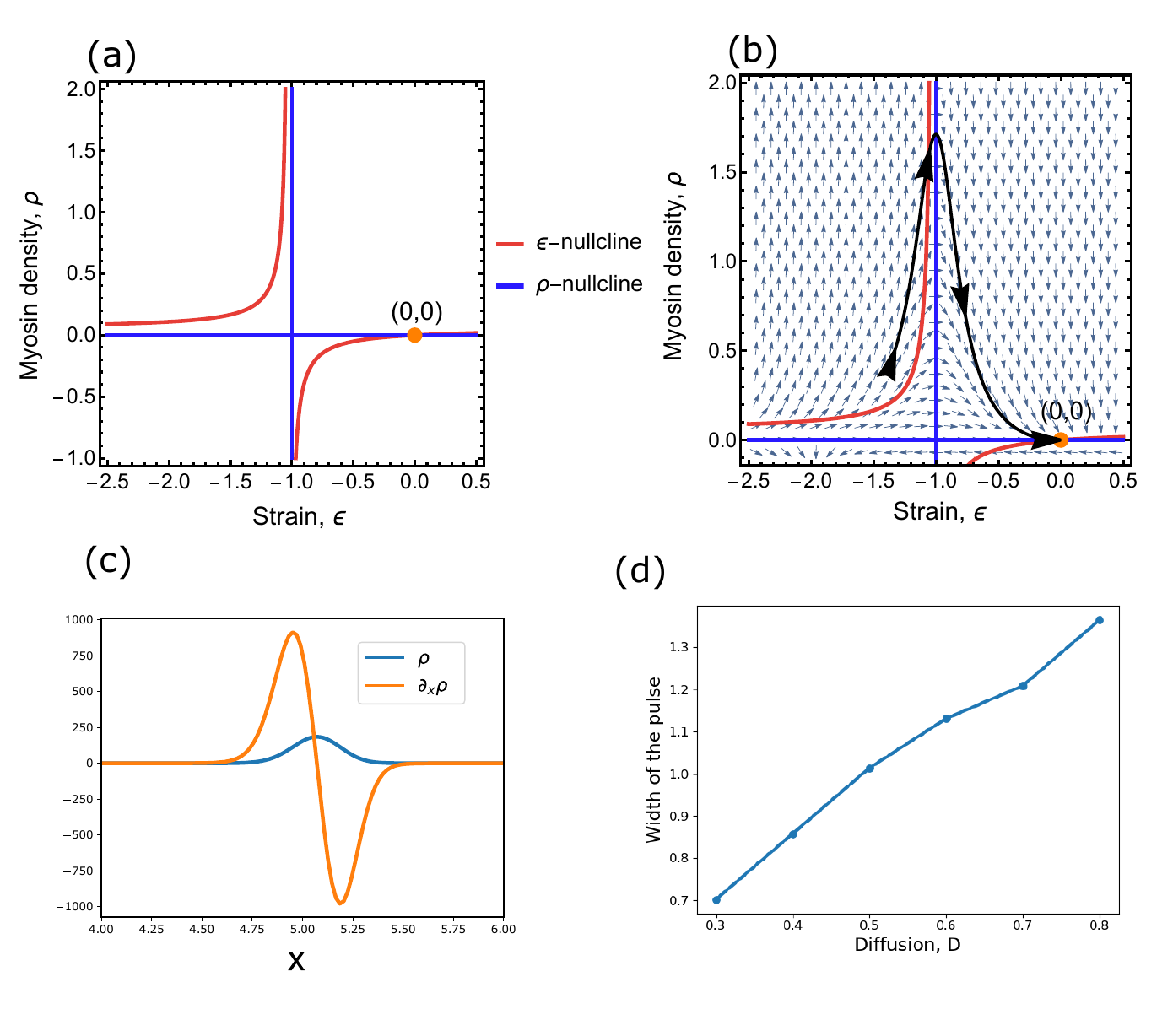} 
\caption[]{(a)Red and blue curves are nullclines of the system. Orange dot indicates the fixed point at ($\epsilon,\rho$) = (0,0). (b) shows stable flow around the fixed point. Black solid line is a trajectory starting sufficiently far from the fixed point showing a large excursion before returning to the fixed point. (c) shows that spatial derivative of $\rho$ (in orange) is many folds larger than $\rho$ (in blue) in a compact region of space which points to the existence of an interior boundary layer. (d) Width of the pulse reduces as diffusion D is reduced showing the presence of a boundary layer as D $\to 0$. Parameter values for (a,b) are set to $ v = -3.9, B-c^2 = 5, \zeta_1 = 9.68, \chi_0 = 1, D = 1, \zeta_2 = 0.2$}
\label{fig:nullclinepulse}
\end{figure}

For simplicity, we drop the 
viscous stress term and integrate the conserved dynamical system once to obtain, 
\begin{eqnarray}
\label{eq:eps_pulsenullcline}
(B-c^2)\,\epsilon' &=&   -\biggl(\frac{-D v  \zeta _2^2 \rho ^2 \epsilon + \left(\zeta _1 v -2 D \zeta _2 v\right) \rho \epsilon- D v \epsilon + \zeta _1 v  \rho}{D \left(\zeta _2 \rho +1\right){}^2}\biggr)
\end{eqnarray}
\begin{eqnarray}
\label{eq:rho_pulsenullcline}
D \,\rho'  &=& v \,\rho + v \,\rho \,\epsilon
\end{eqnarray} 
We see that unlike in the wavetrain case,  the highest order derivative term, $D \rho'$,  cannot be dropped. For as we see from the numerically obtained snapshots (Fig.\,\ref{fig:nullclinepulse}(c)), the profile of $\rho$ in the comoving frame has a boundary layer, where $\rho'$ gets larger as $D \to 0$ and supported by the fact that the pulse width shrinks as D is reduced (Fig.\,\ref{fig:nullclinepulse}(d)). 
Here we analyse the phase portrait of this system, with the $\rho$-nullcline given by
 $ \epsilon=-1$ or $\rho=0$, and the $\epsilon$-nullcline given by 
 $$\epsilon = \frac{\zeta _1 \rho }{D \zeta _2^2 \rho ^2+2 D \zeta _2 \rho + D-\zeta _1 \rho  \chi _0} \, .  $$
As indicated in Fig.\,\ref{fig:nullclinepulse}(a), these  have intersections only at $(\epsilon,\rho) = (0,0)$. Using parameter values that correspond to the travelling pulse phase (Sect.\,\ref{subsec:numerical}), we plot the phase portrait of the dynamical system around this fixed point. Although Fig.\,\ref{fig:nullclinepulse}(b) shows that (0,0) (orange point) is a stable fixed point, trajectories starting from  initial conditions sufficiently away from the fixed point, execute large excursions before returning to it (black curve).  
Such large excursions in phase space are indicative of a homoclinic orbit~\cite{winfree1980geometry}, which we confirm using a more sophisticated boundary-layer analysis. 
\subsubsection{Boundary layer analysis}
\label{subsect:b-layer}
To prove the existence of a homoclinic orbit and obtain its analytical solution, we begin with the following set of odes,
\begin{eqnarray}\label{boundarylayer_eps_1}
\epsilon' &=& \delta  
\end{eqnarray}
\begin{eqnarray}\label{boundarylayer_delta_1}
v \delta' &=&  v  \epsilon - (B-c^2) \delta  - (\frac{\zeta_1 \rho}{(1+\zeta_2\rho)})'
\end{eqnarray}
\begin{eqnarray}\label{boundarylayer_rho_1}
\frac{D}{v} \rho' &=&   \rho(1+ \epsilon)\, ,
\end{eqnarray}
obtained by integrating Eqs.\,\eqref{eq:tpnullclineeps} and \eqref{eq:tpnullclinerho} once and setting $\epsilon'=\delta$. 

\begin{figure}[h]
\centering
\includegraphics[width=10.5cm]{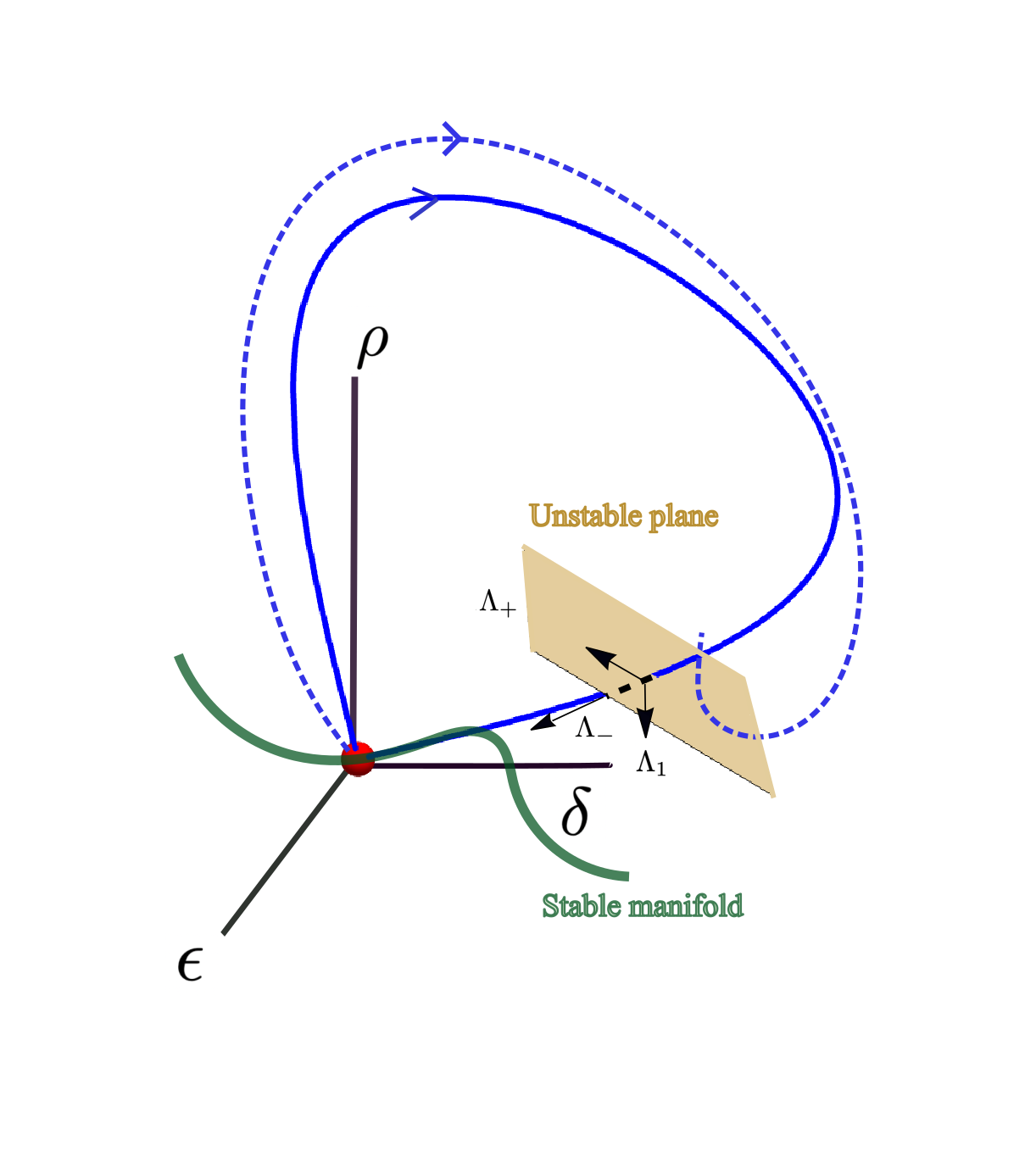}
\caption[]{Schematic representation of the saddle fixed point (red) at the origin in \((\epsilon, \delta, \rho)\) space. The solid blue line illustrates a homoclinic orbit, which starts from the vicinity fixed point, undergoes a large excursion due to nonlinearities, and returns precisely along the stable direction \(\Lambda_-\) into the stable manifold (solid green curve) as \(D/v \to 0\). Any deviation from this condition results in a failed homoclinic orbit (dashed blue line), since the associated flow vector has nonzero components along the unstable directions $\Lambda_1$, $\Lambda_+$ veering off from the fixed point.}
\label{fig:homoclinic}
\end{figure}

It is convenient to separate the linear and nonlinear terms in Eqs.\,\eqref{boundarylayer_eps_1}-\,\eqref{boundarylayer_rho_1}, and write them in matrix form as,
\begin{eqnarray}
    \begin{bmatrix}
        \epsilon' \\
        \delta'\\
        \rho'
    \end{bmatrix} &=&
    \underbrace{\begin{bmatrix}
        0 & 1 & 0\\
        1 & \frac{-(B-c^2)}{v} & \frac{-\zeta_1}{D}\\
        0 & 0 & \frac{v}{D}
    \end{bmatrix}}_{\text{Linear terms}}
    \begin{bmatrix}
        \epsilon \\
        \delta \\
        \rho
    \end{bmatrix}
    + \underbrace{\begin{bmatrix}
        0\\
        -\frac{\zeta_1 \rho \epsilon}{D(1+\zeta_2 \rho)^2}\\
        \frac{v\rho \epsilon}{D}
    \end{bmatrix}}_{\text{Nonlinear terms}}
\label{eq:blayerseplinnonlin}
\end{eqnarray}
The eigenvalues ${\bs \lambda}$ and eigenvectors ${\bs \Lambda}$  of the linear matrix are,
\begin{eqnarray}
    \lambda_1 = \frac{v}{D} \,\,\,\,\,\,\,\,\,\,\,\mbox{corresponding to}\,\,\,\,\,\,\,\,\, \Lambda_1=\begin{bmatrix}
        \frac{D \zeta_1}{D^2 - (B-c^2) D - v^2} \\
        \frac{v \zeta_1}{D^2 - (B-c^2) D - v^2} \\
        1
    \end{bmatrix} \nonumber \\ 
    \lambda_\pm = \frac{-(B-c^2) \pm \sqrt{(B-c^2)^2 + 4v^2}}{2v} \,\,\,\,\,\,\,\,\,\,\,\,\mbox{corresponding to}\,\,\,\,\,\,\,\,\, \Lambda_\pm=\begin{bmatrix}
        \frac{(B-c^2) \pm \sqrt{(B-c^2)^2 + 4v^2}}{2v} \\
        1 \\
        0
    \end{bmatrix}
\end{eqnarray}
We note that $\lambda_1$ and $\lambda_+$ are always positive (unstable directions) and $\lambda_-$ is always negative (stable direction). 
This shows that the fixed point at $(0,0,0)$ is a saddle fixed point (red point in Fig.\,\ref{fig:homoclinic}). A travelling pulse solution in this analysis translates to a homoclinic orbit (solid blue line) which starts from the vicinity of this fixed point and executes a large excursion in phase space before returning to it. We note that such a homoclinic trajectory can only return to the fixed point if the effect of the nonlinearities act in such a way as to make the associated flow direction exactly coincide with the single stable direction, $\Lambda_-$ (see Fig.\,\ref{fig:homoclinic}). Even a slight deviation from this scenario will give rise to a nonzero projection of the flow onto the unstable directions and hence
drive the trajectory away from the fixed point (dashed curve in Fig.\,\ref{fig:homoclinic}).

To show this explicitly, we decompose the position vector along the trajectory in the 3d phase space in  the eigenvector ${\bs \Lambda}$ basis,
\begin{eqnarray}
    \begin{bmatrix}
        \epsilon\\
        \delta\\
        \rho
    \end{bmatrix} = a_1(\xi) \Lambda_1 + a_+(\xi) \Lambda_+ + a_-(\xi) \Lambda_-\, .
\end{eqnarray}

Note that this is not an orthonormal basis, as the linear matrix in Eq.\,\eqref{eq:blayerseplinnonlin} is not symmetric. From this decomposition, it follows that a trajectory will be homoclinic only if, as it approaches the fixed point, both components along the unstable directions ($a_1$ and $a_+$) vanish from the positive side, before the stable component $a_-$ goes to zero. To decide which of the unstable components $a_1$ or $a_+$ vanishes first, we note that all the nonlinear terms in Eq.\,\eqref{eq:blayerseplinnonlin} depend on $\rho$. Now if $a_1$ were to vanish before $a_+$, then $\rho$ would become $0$, forcing  the trajectory to lie on the unstable $(\epsilon, \delta)$ subspace, driving it away from the fixed point.


This imposes a strict hierarchy in the vanishing of the components: first, $a_+ \to 0$, followed by $a_1 \to 0$, and finally $a_- \to 0$ along the stable manifold into the fixed point. The necessary condition to ensure this sequence is that the parameter $D/v$  be small. This can be seen by re-expressing the dynamical equations \eqref{eq:blayerseplinnonlin} in terms of ${\bs a}$, 
\begin{eqnarray}
    a^{\prime}_1 \Lambda_1 + a^{\prime}_+  \Lambda_+ + a^{\prime}_- \Lambda_- &=& a_1 \lambda_1 \Lambda_1 + a_+ \lambda_+ \Lambda_+ + a_- \lambda_- \Lambda_- \nonumber \\ &+& \begin{bmatrix}
        0 \\
       - \frac{\zeta_1 a_1\biggl(a_1\frac{D \zeta_1}{D^2 - (B-c^2) D - v^2} + a_+ \frac{(B-c^2) + \sqrt{(B-c^2)^2+4v^2}}{2v} + a_- \frac{(B-c^2) - \sqrt{(B-c^2)^2+4v^2}}{2v}\biggr)}{D(1+\zeta_2 a_1)^2} \\
       \frac{v a_1\biggl(a_1\frac{D \zeta_1}{D^2 - (B-c^2) D - v^2} + a_+ \frac{(B-c^2) + \sqrt{(B-c^2)^2+4v^2}}{2v} + a_- \frac{(B-c^2) - \sqrt{(B-c^2)^2+4v^2}}{2v}\biggr)}{D}
    \end{bmatrix}\, .
    \label{eq:compdynfullblayer}
\end{eqnarray}
We project the above equation along the unstable directions. Multiplying both sides of the above by $[0,0,1]$, we obtain the dynamics along $\Lambda_1$ 
\begin{eqnarray}
    a_{1}^{\prime} = \frac{v}{D}a_{1} \Biggl[1 + \biggl(a_1 \frac{D \zeta_1}{D^2 - (B-c^2)D - v^2} + a_+ \frac{(B-c^2) + \sqrt{(B-c^2)^2 + 4v^2}}{2v} + a_- \frac{(B-c^2)-\sqrt{(B-c^2)^2 + 4v^2}}{2v}\biggr)\Biggr]\, .
    \nonumber
\end{eqnarray}


Multiplying both sides of Eq.\,\eqref{eq:compdynfullblayer} by $\Lambda_{+}^T$ and simplifying, we obtain dynamics along the other unstable direction $\Lambda_+$, 
\begin{eqnarray}
    a^{\prime}_+ &=& a_+ \lambda_+ - \underbrace{\frac{a_1}{\Lambda_{+}^T \Lambda_+}}_{>0}\underbrace{\Biggl[-\frac{v}{D(\frac{v^2}{D^2}+\frac{(B-c^2)}{D}-1)}(\frac{(B-c^2) + \sqrt{(B-c^2)^2+4v^2}}{2v}+\frac{v}{D}) + \frac{1}{(1+\zeta_2 a_1)^2}\Biggr]}_{< 0 \,\,\,\text{when}\,\,\, \frac{D}{v}\to 0 }\frac{\zeta_1}{D} \nonumber \\ &&\underbrace{\Biggl(a_1 \frac{D \zeta_1}{D^2 - (B-c^2)D - v^2} + a_+ \frac{(B-c^2) + \sqrt{(B-c^2)^2 + 4v^2}}{2v} + a_- \frac{(B-c^2)-\sqrt{(B-c^2)^2 + 4v^2}}{2v}\Biggr)}_{<-1} \, .
    \nonumber
\end{eqnarray}
It now follows that,
\begin{enumerate}
    \item In the dynamical equation for $a_1$, since $\frac{v}{D} a_{1}$ is positive, we see that $a_1$ can go to zero only when the term inside the square brackets is negative, which occurs when the term in round brackets (which represents $\epsilon$ in terms of its components) is less than $-1$.  

    \item Similarly, $a_+$ can become zero only when the sum of the nonlinear terms (inside both the square and round brackets) is negative. Since the term in round brackets is already less than $-1$ (argument (1)), the full nonlinear term will be negative only when the term inside the square brackets is also negative. This condition is generically satisfied when $\frac{D}{v} \to 0$. 
\end{enumerate}
This hierarchical vanishing of components along the unstable directions when $\frac{D}{v} \to 0$, ensures that the trajectory aligns along the stable direction and ultimately lands on the stable manifold, thus completing a homoclinic orbit (solid curve in Fig.\,\ref{fig:homoclinic}). This necessary condition is consistent with the fact that $D$, does not represent the physical diffusion of bound myosin, but was introduced as a numerical regularizer.  


If the ratio \(\Delta = D/v\) is small but nonzero, the system exhibits near homoclinic orbits, the return trajectory stays close to the homoclinic, but then veers away from the fixed point. We may compute the time \(\xi\), taken by this orbit to veer away from a saddle fixed point, to leading order in $\Delta$. This can be analyzed by considering the linearized dynamics near the fixed point \cite{strogatz2018nonlinear} in the unstable $(\epsilon, \delta)$ subspace.  The general solution for \(\epsilon(\xi)\), expressed in terms of the eigenvalues of the linearized system, is
\begin{equation}
    \epsilon(\xi) = a_1 e^{\lambda_- \xi} + b_1 e^{\lambda_+ \xi} + c_1 e^{\lambda_1 \xi},
\end{equation}
where \(\lambda_-\) is the eigenvalue associated with the stable direction, and \(\lambda_+\) and \(\lambda_1\) correspond to two unstable directions. Since we are interested in the time it takes the orbit to depart from the fixed point, the contribution from the stable direction can be neglected. Furthermore, the dynamics has a larger component along the unstable direction associated with the dominant eigenvalue, \(\lambda_+\), thus we approximate
\begin{equation}
    \epsilon(\xi) \approx b_1 e^{\lambda_+ \xi}.
\end{equation}
We obtain $b_1$, by setting $\epsilon$ to be a distance \(\Delta\) from the homoclinic orbit at \(\xi = 0\): this gives \(b_1 = \Delta\).
The escape time is the time it takes for the deviation to be of order $1$, leading to 
\begin{equation}
    \xi = \frac{-\ln \Delta}{\lambda_+}.
\end{equation}
This is related to the lifetime of the travelling pulse, and grows logarithmically with the inverse of the initial deviation and is inversely proportional to the unstable eigenvalue \(\lambda_+\), when $\Delta$ is small but nonzero.

Our arguments indicate that the system of equations \eqref{boundarylayer_eps_1} - \eqref{boundarylayer_rho_1}  has an interior boundary layer of width $D/v$ in $\rho$ \,\cite{strogatzyoutubeblyaer,eggers2015singularities,bender2013advanced}.
\begin{figure}[h]
\centering
\includegraphics[width=14.5cm]{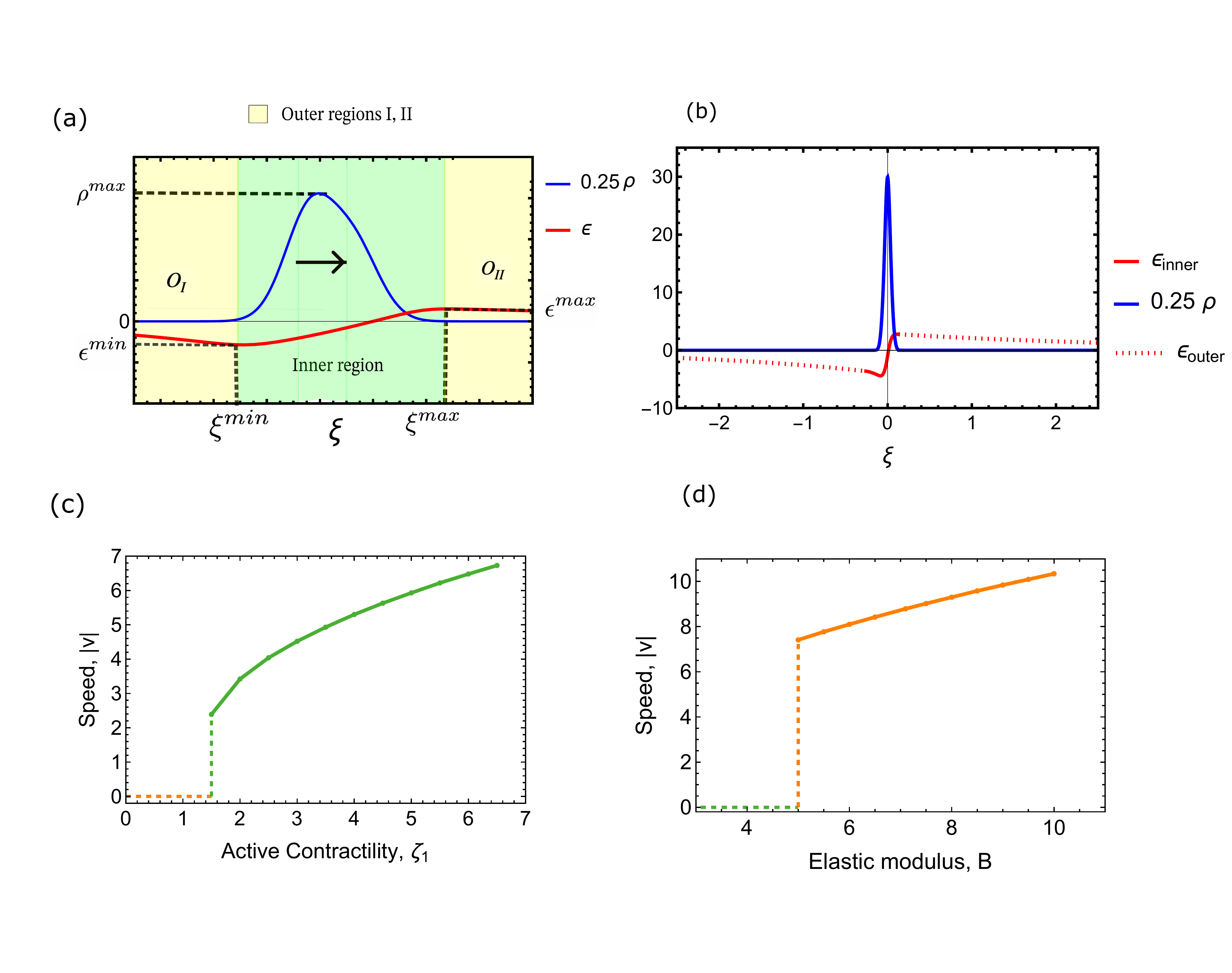}
\caption[]{(a)Schematic showing the method of obtaining an analytical solution using boundary layer analysis (Outer and inner regions are coloured in yellow and green respectively). Matching conditions and constraints of the system are used to obtain the unknowns of the system such as $v$, $\rho^{max},\epsilon^{min},\xi^{min}, \epsilon^{max}, \xi^{max}$. (b) An asymptotically matched analytical solution for $\rho$ (in blue) and a piecewise matched solution for $\epsilon$ (in red) obtained by a least square optimization (see text) with constants $v=-5.3$, $c_2 = 8.03$ , $\epsilon^{min} = -3.61$, $\xi^{min}=0.26$, $\epsilon^{max}=2.77$ $\xi^{max}=0.13$ at $B=5$ and $\zeta_1 = 4$ (c) A plot of speed ($|v|$) versus active contractility ($\zeta_1$) at $B=5$. (d) A plot of speed ($|v|$) versus elastic modulus ($B$) at $\zeta_1 =8$. Rest of the parameters are set to $D=0.5$ and $R=10$.}
\label{fig:boundlayer}
\end{figure}
This allows us to divide the domain into three different regions (Fig.\ref{fig:boundlayer}(b)), inner and outer ($O_{I}$ and $O_{II}$) regions, where the equations get simplified. Since we are interested in travelling pulse solutions, that is, compact profiles which go to zero at the domain boundaries, we look for solutions consistent with these boundary conditions in the inner and outer regions --\\

\noindent
\textbf{Inner region}: Here, $\rho$ rises to a large maximum value ($\rho^{max}$) before falling abruptly (green region in Fig.\ref{fig:boundlayer}(a)). In this interior region, the small parameter $D/|v|$ multiplies $\rho^{\prime}$ which is large, and so cannot be dropped.
We rescale our independent variable in terms the stretched coordinate $X = \frac{\xi}{\Delta}$, where the small parameter $\Delta \equiv \frac{D}{|v|}$. Equations \eqref{boundarylayer_eps_1} and \eqref{boundarylayer_rho_1} transform to 
\begin{eqnarray}
 v \epsilon'' &=& \Delta^2  v  \epsilon -\Delta (B-c^2) \epsilon' - \Delta (\frac{\zeta_1 \rho}{(1+\zeta_2\rho)}\chi_0)'
 \label{eq:epsilon_bl_zo}
\end{eqnarray}
and
\begin{eqnarray}
\rho' &=&   \text{sgn}(v) \rho(1+ \epsilon)\, .
\label{eq:rho_bl_zo}
\end{eqnarray}
which describe the profile in the inner region.\\

\noindent
\textbf{Outer regions} ($O_{I}$ and $O_{II}$): In this region, since $\rho'$ is small, the $\Delta\, \rho'$ term can be ignored. It follows from Eq.\,\eqref{boundarylayer_rho_1} that, $\rho = 0$ or $\epsilon=-1$. We choose $\rho=0$ because it agrees with the desired boundary conditions at $\xi=\pm \infty$. 
This together with the reduced dynamics for $\epsilon$ from Eqs.\,\eqref{boundarylayer_eps_1}-\eqref{boundarylayer_delta_1}, 
\begin{eqnarray} \label{boundarylayer_eps_3}
v \epsilon'' &=& v \epsilon- (B-c^2) \epsilon'
 \label{eq:epsouter}
\end{eqnarray}
describes the profile in the outer regions. 
\\

Having described the dynamical equations in the inner and outer regions, we set up a perturbation scheme by expanding $\rho$ and $\epsilon$ in terms of a power series in the small parameter $\Delta$, 
\begin{eqnarray}
\label{eq:serexp}
    \rho &=& \rho_0 + \Delta \rho_1 + \Delta^2 \rho_2 + \ldots \\ \nonumber
    \epsilon &=& \epsilon_0 + \Delta \epsilon_1 + \Delta^2 \epsilon_2 + \ldots
\end{eqnarray}
and demanding smoothness and continuity of the variables across the inner and outer regions. That is, we plug the series expansion \eqref{eq:serexp} in the equations describing the profiles in the outer and inner regions and solve for the variables ($\rho_n, \epsilon_n$) order by order in $\Delta$. We then match the function and its derivatives across the regions to all orders in $\Delta$. The matching conditions can be solved to obtain the full profile upto a given order in $\Delta$ (Fig.\,\ref{fig:boundlayer}(a)). \\

(i) \underline{Zeroth order}: From Eq.\,\eqref{eq:serexp},
at zeroth order in $\Delta$, $\rho=\rho_0$ and $\epsilon=\epsilon_0$, and the system of equations in the inner region Eqs.\,\eqref{eq:epsilon_bl_zo},\,\eqref{eq:rho_bl_zo} reduce to, 
\begin{eqnarray} \label{boundarylayer_eps_2}
v \epsilon_0'' &=& 0
\end{eqnarray}
\begin{eqnarray}\label{boundarylayer_rho_2}
\rho_0' &=&   \text{sgn}(v)\rho_0(1 + \epsilon_0)
\end{eqnarray}
Solving Eq.\,\eqref{boundarylayer_eps_2}, we have, 
\begin{eqnarray}
\epsilon_0 &=& c_1 + X c_2
\label{eq:eps_bl_gensol}
\end{eqnarray}
Substituting this in Eq.\,\eqref{boundarylayer_rho_2} we obtain,
\begin{eqnarray}
\rho_0' &=&   \text{sgn}(v)\rho_0(1 + c_1 + c_2 X) \, .
\label{boundarylayer_rho_2.5}
\end{eqnarray}
The inner solution to zeroth order is given by
\begin{eqnarray}\label{boundarylayer_rho_3}
\rho^{inner}_0 &=& c_3 e^{\text{sgn}(v)(X + c_1 X + \frac{c_2 X^2}{2})}
\end{eqnarray}
where $v$, $c_1$, $c_2$ and $c_3$ are constants that need to be determined by matching with the outer solution $\rho^{outer}$ and $\epsilon^{outer}$ and using other constraints.

The inner and outer solutions of $\rho$ are  matched asymptotically,
\begin{eqnarray}
    \lim_{X\to\pm\infty} \rho^{inner}_0 (X) = \lim_{\xi\to 0} \rho^{outer}(\xi)
\end{eqnarray}
Since $\rho^{outer} = 0$, we obtain conditions for the constants, $c_1 = -1$ and the sign of $c_2$ is set by $sgn(v)$. Henceforth, for specificity, we consider a right moving pulse, $v<0$ and so $c_2>0$. As we have noted, for every travelling solution moving to the right with velocity $-v$, there exists a parity transformed solution moving to the left with velocity $+v$.
Since the product $\Delta\,\rho' = 0$ in the outer region, this asymptotic matching condition for $\rho$ 
holds at all orders in $\Delta$. 

Now substituting this value of $c_1$ in Eq.\,\eqref{eq:eps_bl_gensol}, we have, 
\begin{eqnarray}
\epsilon[X=0] = c_1 = -1   \, ,
\end{eqnarray}
which together with Eqs.\,\eqref{boundarylayer_rho_1} and \eqref{boundarylayer_rho_2.5} suggests that $\rho'=0$ and $\rho''<0$ at $X=0$, a maximum (Fig.\,\ref{fig:boundlayer}(a)).


To find $c_3$, we use the fact that the dynamics of $\rho$ is a conservation equation (Eq.\,\eqref{Eq:nucleation4}), implying that the total myosin density in the pulse, i.e., $\int_{-\infty}^{\infty} \rho d\xi = R$ is fixed. Using the zeroth order solution for $\rho$ in this condition, we have
\begin{eqnarray}
    \int_{-\infty}^{\infty} c_3 e^{\frac{-c_2}{2} (\frac{\xi v}{D})^2} d\xi= R < \infty
\end{eqnarray}
Note that this integral will converge to a finite value only if $c_2>0$ and the magnitude of $c_2 \sim O(\frac{D^2}{v^2})$, leading to 
\begin{eqnarray}
   c_3 = R \sqrt{\frac{c_2 v^2}{2 \pi D^2}}\, .
\end{eqnarray}
Note that to this order, the density profile is symmetric about the origin, while the strain profile exhibits an asymmetry. The former is contrary to our demonstration of an asymmetric moving profile in Sect.\,\ref{sect:nuctrans}

We now look at the solution for the strain field (Eq.\,\eqref{eq:eps_bl_gensol}) in the $\xi$ coordinate.
\begin{eqnarray}
\epsilon_0 = -1 + c_2 \frac{|v|}{D} \xi \label{eq:strainzeroxi}
\end{eqnarray}
Since $c_2>0$, the slope of this profile is positive (Fig.\,\ref{fig:boundlayer} (a)).

We now match this inner solution of $\epsilon$ to its outer solution. 
The outer solutions for $\epsilon$ are the asymptotic solutions of the linear equation\,\eqref{boundarylayer_eps_3} satisfying the boundary conditions $\epsilon = 0$ at $\xi = \pm \infty$, 

\noindent
\begin{eqnarray}
\epsilon^{outer} &=& \epsilon^{min} e^{\lambda_1 (\xi+\xi^{min})} \,\,\,\text{for} \,\,\, \xi < -\xi^{min} \,\,\,(\text{Region}\,\,\, O_{I}) \\ \nonumber
\epsilon^{outer} &=& \epsilon^{max} e^{\lambda_2 (\xi-\xi^{max})} \,\,\,\text{for} \,\,\, \xi > \xi^{max} \,\,\,(\text{Region}\,\,\, O_{II})
\end{eqnarray}
where $\lambda_1$ ($\lambda_2$) is positive (negative). Substituting this in Eq. \ref{boundarylayer_eps_3}, we find the inverse decay lengths, 
\begin{eqnarray}
    \lambda_{1,2} = \frac{-(B-c^2)\mp\sqrt{(B-c^2)^2 + 4v^2}}{2 v} \,.
\end{eqnarray}

With this, the constants to be determined are $c_2$, $v$, $\xi^{min}$, $\xi^{max}$, $\epsilon^{min}$ and $\epsilon^{max}$. We impose smoothness and continuity of the strain field $\epsilon$, at the boundary of the inner and outer regions. At zeroth order, we see that the derivative of the inner solution (Eq.\,\eqref{eq:strainzeroxi}), $c_2 |v|/D$ is positive for the right moving pulse and can never be matched with the outer solution, $\lambda_2$ which is negative. This results in an underdetermined system and encourages us to explore higher order corrections of $\rho$ and $\epsilon$ in the inner region. \\

(ii) \underline{First order}: 
From Eq.\,\eqref{eq:serexp},
at first order in $\Delta$, $\rho=\rho_0 + \Delta \rho_1$ and $\epsilon=\epsilon_0 + \Delta \epsilon_1$, and the system of equations in the inner region are,

\begin{eqnarray}
    v \epsilon_1' &=& -(B-c^2) \epsilon_0 - \frac{\zeta_1 \rho_0}{1 + \zeta_2 \rho_0}
    \label{eq:firstordereps}
\end{eqnarray}
\begin{eqnarray}
    \rho_1' = -(\rho_1 + \rho_0 \epsilon_1 + \epsilon_0\rho_1)
     \label{eq:firstorderrho}
\end{eqnarray}

While the $\epsilon$ profile was shown to be asymmetric at the zeroth order, we see from the above equations, that the asymmetry in the $\rho$ profile  appears at first order.
For while the first term in the RHS of Eq.\,\eqref{eq:firstordereps} is odd, the second is even ($\frac{\zeta_1 \rho_0}{1+ \zeta_2 \rho_0}$ term), which implies that $\epsilon_1$ and hence $\rho_1$ (Eq.\,\eqref{eq:firstorderrho}) is in general asymmetric (see Eq.\eqref{eq:rho1solnbl} for form of $\rho_1$). 


However the mismatch of slopes of inner and outer solutions of $\epsilon$ still remains. This can be deduced from Eq.\,\eqref{eq:firstordereps} by noting that for the right moving pulse, $\epsilon_1'$ can never be negative in the $\xi>0$ region because both $\epsilon_0>0$, $\rho_0>0$ and $v<0$ (and $B-c^2 >0$, mechanical stability). 
Thus $\epsilon' = \epsilon_0' + \Delta \epsilon_1'$ can never become negative and hence can never match the negative slope of the exponential tail ($\lambda_2<0$). 
We are forced to include a second order correction. \\

\noindent
(iii) \underline{Second order}: From Eq.\,\eqref{eq:serexp},
at second order in $\Delta$, $\rho=\rho_0 + \Delta \rho_1 + \Delta^2 \rho_2$ and $\epsilon=\epsilon_0 + \Delta \epsilon_1 + \Delta^2 \epsilon_2$, and the system of equations in the inner region are, 
\begin{eqnarray}
     \epsilon_2'' &=& \epsilon_0 - \frac{(B-c^2)}{v} \epsilon_1' - \frac{\zeta_1}{v} \rho_1'
\label{eq:secondordereps}
\end{eqnarray}
\begin{eqnarray}
    \rho_2' &=& -(\rho_2 + \rho_1 \epsilon_1 + \rho_0 \epsilon_2 + \rho_2 \epsilon_0) \,,
\end{eqnarray}
where for simplicity we have taken the active stress to be linear in $\rho$ (though this is not essential). 
These equations can be solved and the solutions ($\epsilon_n$, $\rho_n$) at orders ($n=0,1,2$) are displayed in Appendix\,\ref{app:boundlayer}. 
For the derivatives to match across the inner and outer regions, it is necessary for $\epsilon_2'$ to be negative at the matching point. 
To determine this, we integrate Eq.\,\eqref{eq:secondordereps} across the boundary layer adjoining $O_{II}$ from $\xi_-$ to $\xi_+$, recognizing that the third term on the RHS of  Eq.\,\eqref{eq:secondordereps} is large and negative in the limit $\Delta \to 0$ (Fig.\,\ref{fig:boundlayer}(a)) and dominates over the first two terms, thus
\begin{eqnarray}
    [\epsilon_2']_{\xi^-}^{\xi_+} &=& -\frac{\zeta_1}{v} [\rho_1]_{\xi^-}^{\xi_+} \,,
\end{eqnarray}
demonstrating that $\epsilon_2'$ is indeed negative in this region. 
A similar argument can be constructed to obtain solutions for the strain field in the $\xi<0$ region (matching with $O_{I}$). 
Matching the strain and its derivatives at $\xi^{min}$ and $\xi^{max}$, 
\begin{enumerate}
    \item $\epsilon_{inner}\mid_{\xi = -\xi^{min}} = \epsilon^{min} e^{\lambda_1 (\xi+\xi^{min})}\mid_{\xi = -\xi^{min}}$
    \item  $\epsilon_{inner}\mid_{\xi = \xi^{max}} = \epsilon^{max} e^{\lambda_2 (\xi-\xi^{max})}\mid_{\xi = \xi^{max}}$
    \item $\epsilon'_{inner}\mid_{\xi = -\xi^{min}} = \lambda_1 \epsilon^{min}$
    \item  $\epsilon'_{inner}\mid_{\xi = \xi^{max}} = \lambda_2 \epsilon^{max} $
    \item  $\epsilon''_{inner}\mid_{\xi = \xi^{max}} = (\lambda_2)^2 \epsilon^{max} $
\end{enumerate}
together with the constraint $\int_{-\infty}^{\infty} \epsilon \,d\xi = 0$ (since $\epsilon\equiv u^{\prime}$), gives 6 conditions which allows us to evaluate the 6 unknown constants $c_2, v, \epsilon^{max}, \xi^{max}, \epsilon^{min}, \xi^{min}$ in terms of the parameters in the dynamical equations. Since the dependence on these parameters is implicit and nonlinear, we evaluate the constants using a least square optimization. This allows us to build an analytical solution for $\epsilon$ and $\rho$ across the domain (Fig.\,\ref{fig:boundlayer}(b)) and shows good agreement with the numerically obtained profile of the travelling pulse (Fig.\,\ref{fig:numphasdiag}(c)). This agreement may be improved by taking higher order corrections in $\Delta$.  This completes our boundary layer analysis that provides an accurate profile of the myosin density and strain field associated with the travelling pulse moving with an accurately determined velocity $v$.

\section{Spatiotemporal chaos\label{sect:stchaos}} 
Our earlier two-mode analysis in Sect.\,\ref{sect:twomode} showed that the amplitude equations showed temporal chaos in a small regime of contractility. In this section we ask whether the pdes \eqref{eq:epsnum_1d} and\,\eqref{eq:rhonum_1d}, exhibit spatiotemporal chaos in a parameter regime.

\begin {figure}[htp]
\includegraphics[width=12.3cm]{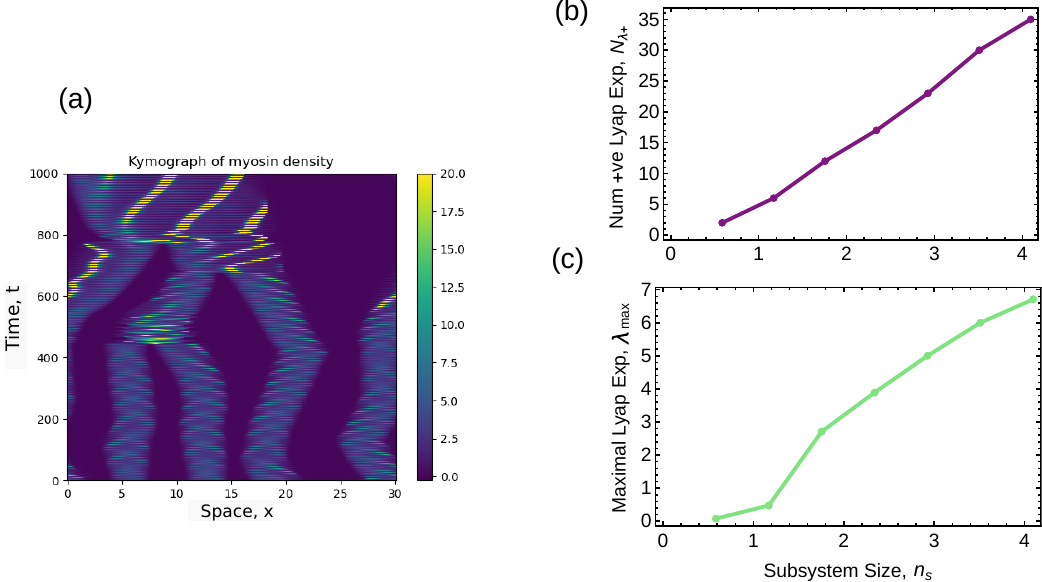}
\caption{(a) Kymograph showing aperiodic solution of myosin density in space-time. (b) Number of positive Lyapunov exponents and (c) Maximal Lyapunov exponent scales with subsystem size. Parameters are set at $B = 2$, $\zeta_1 = 4.25$, $c=0.1,\chi_0=\chi_1=1,\chi _2=0.01,\chi_3=-0.001,D=1,\zeta_2 = 0.215,k=0.2,\alpha =3$}
\label{fig:kymchaos}
\end{figure}

In a small region of the phase diagram between the segregated and the travelling phase, the time series of both myosin density $\rho$ and strain $\epsilon$ is aperiodic (Movie\ref{app:numerical}2). 
To analyse the nature of these aperiodic solutions, we look at the space-time plot (kymograph in Fig.\,\ref{fig:kymchaos}(a)). 
While temporal chaos is revealed by the presence of a few positive Lyapunov exponents, spatiotemporal chaos requires the computation of the full Lyapunov spectrum. A commonly  used method \cite{carretero1999scaling} is to split the full spatial domain into subsystems of increasing sizes. We further reduce these subsystems (multiple spatial points) as multiple time series for which Lyapunov exponents can be obtained using the TISEAN package for nonlinear time series analysis \cite{hegger1999practical}. Lyapunov spectrum of the different subsystems throws up two interesting trends. We see that both number of positive Lyapunov exponents and the maximal Lyapunov exponents scale with subsystem size (Fig.\,\ref{fig:kymchaos}(b)). This feature is known to be a more conclusive validation of spatiotemporal  chaos \cite{das2005routes,chakrabarti2004spatiotemporal}. 

\section{Discussion}
In this paper, we examine  how active contractility and material renewability combine to give rise to mechanically driven excitability in {\it renewable active matter}~\cite{roychowdhury2024active,roychowdhury2024segregation,ankitmadan}. A well studied example is the actomyosin cytoskeleton in cells, either as part of a tissue, or plated on suitably prepared substrates, or even cell extracts. Another biological example is an epithelial tissue where cells undergo proliferation and death. 

To this end we construct a hydrodynamic description of an active elastomer with turnover of its active components (e.g., myosin). Myosin filaments bind to and unbind from the connected actin meshwork, a process that is governed by a phosphorylation cycle.  Once bound, myosin applies local contractile stresses when fueled by an ATP cycle, resulting in a local deformation of the actin meshwork. The myosin unbinding rate responds to this local strain -- in general we consider a catch or slip bond response.  Since the myosin driven force and reaction are controlled by independent chemical cycles, the 
hydrodynamics of the active elastomer is described by a {\it nonreciprocal active field theory}.

Numerical analysis reveals travelling states  among other interesting phases such as a novel segregation \cite{roychowdhury2024segregation}, coexistence (of both static and moving phases), and  spatiotemporal chaos. The travelling states include travelling wavetrains and travelling pulses, which are characteristic of excitable systems, such as 
the Fitzhugh-Nagumo \cite{fitzhugh1961impulses,nagumo1962active} and the slime-mold aggregation \cite{keller1970initiation} models. We perform a phase space analysis after transforming the pdes to a comoving frame, which reveals the existence of these travelling states.
We then study the nucleation and translational instabilities of solitary myosin pulse as a function of increasing contractility.
This analysis provides an analytical estimate of the velocity of the travelling pulse. In the comoving frame, the equations exhibit an interior boundary layer, which we exploit to both prove the existence of a homoclinic orbit and obtain an analytical solution of the myosin profile by asymptotic matching. In doing so, we go beyond the standard linear analysis and provide new mathematical techniques to analyse the nonlinear behaviour of  nonreciprocal active field theories.  

We emphasize that the mechanical excitability described here features active mechanics and material turnover as key players. This is in contrast to the usual excitability contingent on having many system-spanning diffusible species~\cite{winfree1980geometry,keener2010mathematical}. 
It would be interesting to extend  our analysis of mechanical excitability of the renewable cytoskeleton to 2-dimensions,  where a comparison with experimental observations of spiral waves \cite{bement2015activator} can be made.



Our study suggests, that by cell state dependent turnover of crosslinkers and myosin, the {\it in vivo} cytoskeleton can navigate through the space of parameters to achieve a variety of functional phenotypes, including excitability. Although the motivation for our analysis here is mechanical excitability at the scale of the cell, it is clear that this analysis can be extended to other biological systems such as excitable dynamics in epithelial tissues \cite{ankitmadan} and ecosystem dynamics, and indeed to other nonreciprocal active field theories, which generically exhibit a variety of distinct phenotypes -- segregation and coarsening, pattern-formation, mechanical fragility and excitability -- as the parameters of the equations are varied. 


\section{Acknowledgments}
We thank Ayan Roychowdhury and Amit Kumar for useful discussions.
We acknowledge support from the Department of Atomic Energy (India) under project no.\,RTI4006, the Simons Foundation (Grant No.\,287975) and the computational facilities at NCBS.
MR acknowledges Department of Science \& Technology, India for a JC Bose Fellowship (JCB/2018/00030). 

\clearpage
\onecolumngrid
\appendix
\section{Two mode equations arising from Galerkin Mode Truncation}
The dynamical system for two mode Galerkin analysis of Sect.\,\ref{sect:twomode} is presented here. 
\label{app:galerkin}
\begin{eqnarray}
  \dot{\epsilon_1} &=& -\frac{\left(4 \pi ^2 \left(\left(B-c^2\right) k^2+c \left(-k+\zeta _2\right) \zeta _1 \chi_1\right)\right) \epsilon _1}{k^2} \nonumber \\ &+& \frac{\left(c^3 \pi ^2 \left(k-\zeta _2\right) \zeta_1 \chi _3\right) \epsilon _1^3}{2 k^2}-\frac{\left(2 c^2 \pi ^2 \left(k-\zeta _2\right)\zeta _1 \chi _2\right) \epsilon _1 \epsilon _2}{k^2}+\frac{\left(c^3 \pi ^2\left(k-\zeta _2\right) \zeta _1 \chi _3\right) \epsilon _1 \epsilon
   _2^2}{k^2}-\frac{\left(4 \pi ^2 \left(k-2 \zeta _2\right) \zeta _1 \chi _0\right) \rho_1}{k}\nonumber \\ &-&\frac{\left(3 c^2 \pi ^2 \left(k-2 \zeta _2\right) \zeta _1 \chi
   _2\right) \epsilon _1^2 \rho_1}{2 k}+\frac{\left(2 c \pi ^2 \left(k-2 \zeta
   _2\right) \zeta _1 \chi _1\right) \epsilon _2 \rho_1}{k}+\frac{\left(c^3 \pi
   ^2 \left(k-2 \zeta _2\right) \zeta _1 \chi _3\right) \epsilon _1^2 \epsilon _2 \rho_1}{k}\nonumber \\ &-&\frac{\left(c^2 \pi ^2 \left(k-2 \zeta _2\right) \zeta _1 \chi
   _2\right) \epsilon _2^2 \rho_1}{k}+\frac{\left(c^3 \pi ^2 \left(k-2 \zeta
   _2\right) \zeta _1 \chi _3\right) \epsilon _2^3 \rho_1}{4 k}-\left(3 c \pi ^2
   \zeta _2 \zeta _1 \chi _1\right) \epsilon _1 \rho_1^2-\frac{1}{12} \left(5
   c^3 \pi ^2 \zeta _2 \zeta _1 \chi _3\right) \epsilon _1^3 \rho_1^2\nonumber \\ &+&\left(2 c^2 \pi ^2 \zeta _2 \zeta _1 \chi _2\right) \epsilon _1 \epsilon _2 \rho_1^2-\frac{1}{8} \left(7 c^3 \pi ^2 \zeta _2 \zeta _1 \chi _3\right) \epsilon
   _1 \epsilon _2^2 \rho_1^2+\frac{\left(2 c \pi ^2 \left(k-2 \zeta _2\right)
   \zeta _1 \chi _1\right) \epsilon _1 \rho_2}{k}\nonumber \\ &+&\frac{\left(c^3 \pi ^2
   \left(k-2 \zeta _2\right) \zeta _1 \chi _3\right) \epsilon _1^3 \rho_2}{3
   k}-\frac{\left(2 c^2 \pi ^2 \left(k-2 \zeta _2\right) \zeta _1 \chi _2\right) \epsilon
   _1 \epsilon _2 \rho_2}{k}+\frac{\left(3 c^3 \pi ^2 \left(k-2 \zeta _2\right)
   \zeta _1 \chi _3\right) \epsilon _1 \epsilon _2^2 \rho_2}{4 k}+\left(4 \pi ^2
   \zeta _2 \zeta _1 \chi _0\right) \rho_1 \rho_2\nonumber \\ &+&\left(2 c^2 \pi ^2 \zeta _2 \zeta _1 \chi _2\right) \epsilon _1^2 \rho_1 \rho_2-\left(4 c \pi ^2 \zeta _2 \zeta _1 \chi _1\right) \epsilon _2 \rho_1 \rho_2-\frac{1}{4} \left(7 c^3 \pi ^2 \zeta _2 \zeta _1 \chi
   _3\right) \epsilon _1^2 \epsilon _2 \rho_1 \rho_2\nonumber \\ &+&\frac{1}{2} \left(3 c^2 \pi ^2 \zeta _2 \zeta _1 \chi _2\right) \epsilon _2^2 \rho_1 \rho_2-\frac{1}{2} \left(c^3 \pi ^2 \zeta _2 \zeta _1 \chi _3\right) \epsilon _2^3
   \rho_1 \rho_2-\left(2 c \pi ^2 \zeta _2 \zeta _1 \chi _1\right)
   \epsilon _1 \rho_2^2-\frac{1}{24} \left(7 c^3 \pi ^2 \zeta _2 \zeta _1 \chi
   _3\right) \epsilon _1^3 \rho_2^2\nonumber \\ &+&\frac{1}{2} \left(3 c^2 \pi ^2 \zeta _2 \zeta _1 \chi _2 \epsilon _1 \epsilon _2\right) \rho_2^2-\frac{1}{4} \left(3
   c^3 \pi ^2 \zeta _2 \zeta _1 \chi _3\right) \epsilon _1 \epsilon _2^2 \rho_2^2
\label{eq:two_mode_eps1}
\end{eqnarray}
\begin{eqnarray}
    \dot{\epsilon_2} &=& -\frac{\left(4 c^2 \pi ^2 \left(k-\zeta _2\right) \zeta _1 \chi _2\right) \epsilon_1^2}{k^2}-\frac{\left(16 \pi ^2 \left(\left(B-c^2\right) k^2+c \left(-k+\zeta _2\right)
   \zeta _1 \chi _1\right)\right) \epsilon _2}{k^2}\nonumber \\ &+&\frac{\left(4 c^3 \pi ^2 \left(k-\zeta_2\right) \zeta _1 \chi _3\right) \epsilon _1^2 \epsilon _2}{k^2}+\frac{\left(2 c^3 \pi
   ^2 \left(k-\zeta _2\right) \zeta _1 \chi _3\right) \epsilon _2^3}{k^2}+\frac{\left(8 c
   \pi ^2 \left(k-2 \zeta _2\right) \zeta _1 \chi _1\right) \epsilon _1 \rho_1}{k}\nonumber \\ &+&\frac{\left(4 c^3 \pi ^2 \left(k-2 \zeta _2\right) \zeta _1 \chi
   _3\right) \epsilon _1^3 \rho_1}{3 k}-\frac{\left(8 c^2 \pi ^2 \left(k-2 \zeta
   _2\right) \zeta _1 \chi _2\right) \epsilon _1 \epsilon _2 \rho_1}{k}+\frac{\left(3 c^3 \pi ^2 \left(k-2 \zeta _2\right) \zeta _1 \chi
   _3\right) \epsilon _1 \epsilon _2^2 \rho_1}{k}\nonumber \\ &+&\left(8 \pi ^2 \zeta _2 \zeta_1 \chi _0\right) \rho_1^2+\left(4 c^2 \pi ^2 \zeta _2 \zeta _1 \chi
   _2\right) \epsilon _1^2 \rho_1^2-\left(8 c \pi ^2 \zeta _2 \zeta _1 \chi
   _1\right) \epsilon _2 \rho_1^2-\frac{1}{2} \left(7 c^3 \pi ^2 \zeta _2 \zeta
   _1 \chi _3\right) \epsilon _1^2 \epsilon _2 \rho_1^2\nonumber \\ &+&\left(3 c^2 \pi ^2 \zeta_2 \zeta _1 \chi _2\right) \epsilon _2^2 \rho_1^2-\left(c^3 \pi ^2 \zeta _2
   \zeta _1 \chi _3\right) \epsilon _2^3 \rho_1^2-\frac{\left(16 \pi ^2
   \left(k-2 \zeta _2\right) \zeta _1 \chi _0\right) \rho_2}{k}-\frac{\left(4
   c^2 \pi ^2 \left(k-2 \zeta _2\right) \zeta _1 \chi _2\right) \epsilon _1^2 \rho_2}{k}\nonumber \\ &+&\frac{\left(3 c^3 \pi ^2 \left(k-2 \zeta _2\right) \zeta _1 \chi
   _3\right) \epsilon _1^2 \epsilon _2 \rho_2}{k}-\frac{\left(6 c^2 \pi ^2
   \left(k-2 \zeta _2\right) \zeta _1 \chi _2\right) \epsilon _2^2 \rho_2}{k}-\left(16 c \pi ^2 \zeta _2 \zeta _1 \chi _1\right) \epsilon _1 \rho_1 \rho_2\nonumber \\ &-&\frac{1}{3} \left(7 c^3 \pi ^2 \zeta _2 \zeta _1 \chi_3\right) \epsilon _1^3 \rho_1 \rho_2+\left(12 c^2 \pi ^2 \zeta _2
   \zeta _1 \chi _2\right) \epsilon _1 \epsilon _2 \rho_1 \rho_2-\left(6 c^3 \pi ^2 \zeta _2 \zeta _1 \chi _3\right) \epsilon _1 \epsilon
   _2^2 \rho_1 \rho_2+\left(3 c^2 \pi ^2 \zeta _2 \zeta _1 \chi
   _2\right) \epsilon _1^2 \rho_2^2\nonumber \\ &-&\left(12 c \pi ^2 \zeta _2 \zeta _1 \chi_1\right) \epsilon _2 \rho_2^2-\left(3 c^3 \pi ^2 \zeta _2 \zeta _1 \chi
   _3\right) \epsilon _1^2 \epsilon _2 \rho_2^2-\frac{1}{3} \left(5 c^3 \pi ^2
   \zeta _2 \zeta _1 \chi _3\right) \epsilon _2^3 \rho_2^2
\end{eqnarray}
\begin{eqnarray}
  \dot{\rho_1} &=& \left(\frac{4 B \pi ^2}{k}-\frac{4 c^2 \pi ^2}{k}-\alpha  (1+\alpha )-\frac{c \left(k^3+4k \pi ^2 \zeta _1 \chi _1-4 \pi ^2 \zeta _2 \zeta _1 \chi _1\right)}{k^3}\right)
   \epsilon _1+\frac{\left(c^3 \pi ^2 \left(-k+\zeta _2\right) \zeta _1 \chi _3\right)
   \epsilon _1^3}{2 k^3} \nonumber \\ &+& \frac{\left(2 c^2 \pi ^2 \left(k-\zeta _2\right) \zeta _1 \chi_2\right) \epsilon _1 \epsilon _2}{k^3}-\frac{\left(c^3 \pi ^2 \left(k-\zeta _2\right)
   \zeta _1 \chi _3\right) \epsilon _1 \epsilon _2^2}{k^3}+\left(-k-4 d \pi ^2+\frac{4
   \pi ^2 \zeta _1 \chi _0}{k}-\frac{8 \pi ^2 \zeta _2 \zeta _1 \chi _0}{k^2}\right) \rho_1\nonumber \\ &+&\frac{\left(c^2 \pi ^2 \left(5 k-8 \zeta _2\right) \zeta _1 \chi
   _2\right) \epsilon _1^2 \rho_1}{2 k^2}+\left(4 B \pi ^2-\frac{1}{2} k
   \alpha  (1+\alpha )-\frac{2 c \pi ^2 \left(2 c k^2+\left(3 k-4 \zeta _2\right) \zeta
   _1 \chi _1\right)}{k^2}\right) \epsilon _2 \rho_1\nonumber\\ &-& \frac{\left(c^3 \pi ^2 \left(2 k-3 \zeta _2\right) \zeta _1 \chi _3\right) \epsilon _1^2 \epsilon _2 \rho_1}{k^2} +\frac{\left(c^2 \pi ^2 \left(k-2 \zeta _2\right) \zeta _1 \chi_2\right) \epsilon _2^2 \rho_1}{k^2}-\frac{\left(c^3 \pi ^2 \left(3 k-4
   \zeta _2\right) \zeta _1 \chi _3\right) \epsilon _2^3 \rho_1}{4
   k^2}\nonumber \\ &+& \frac{\left(c \pi ^2 \left(-2 k+7 \zeta _2\right) \zeta _1 \chi _1\right)
   \epsilon _1 \rho_1^2}{k}+\frac{\left(c^3 \pi ^2 \left(-4 k+13 \zeta
   _2\right) \zeta _1 \chi _3\right) \epsilon _1^3 \rho_1^2}{12
   k}+\frac{\left(2 c^2 \pi ^2 \left(k-3 \zeta _2\right) \zeta _1 \chi _2\right) \epsilon
   _1 \epsilon _2 \rho_1^2}{k}\nonumber \\ &+&\frac{\left(c^3 \pi ^2 \left(-6 k+19 \zeta
   _2\right) \zeta _1 \chi _3\right) \epsilon _1 \epsilon _2^2 \rho_1^2}{8
   k}\nonumber \\ &-&\left(2 \pi ^2 \zeta _2 \zeta _1 \chi _0\right) \rho_1^3-\left(c^2 \pi^2 \zeta _2 \zeta _1 \chi _2\right) \epsilon _1^2 \rho_1^3+\left(2 c \pi ^2
   \zeta _2 \zeta _1 \chi _1\right) \epsilon _2 \rho_1^3\nonumber \\ &+&\frac{1}{8} \left(7c^3 \pi ^2 \zeta _2 \zeta _1 \chi _3\right) \epsilon _1^2 \epsilon _2 \rho_1^3-\frac{1}{4} \left(3 c^2 \pi ^2 \zeta _2 \zeta _1 \chi _2\right)
   \epsilon _2^2 \rho_1^3+\frac{1}{4} \left(c^3 \pi ^2 \zeta _2 \zeta _1 \chi
   _3\right) \epsilon _2^3 \rho_1^3\nonumber \\ &+&\left(-2 B \pi ^2-\frac{1}{2} k \alpha (1+\alpha )+\frac{2 c \pi ^2 \left(c k^2+\zeta _2 \zeta _1 \chi _1\right)}{k^2}\right)
   \epsilon _1 \rho_2\nonumber \\ &-&\frac{\left(c^3 \pi ^2 \left(k-2 \zeta _2\right) \zeta_1 \chi _3\right) \epsilon _1^3 \rho_2}{3 k^2}+\frac{\left(2 c^2 \pi ^2
   \left(2 k-3 \zeta _2\right) \zeta _1 \chi _2\right) \epsilon _1 \epsilon _2 \rho_2}{k^2}+\frac{\left(c^3 \pi ^2 \left(-4 k+7 \zeta _2\right) \zeta _1 \chi
   _3\right) \epsilon _1 \epsilon _2^2 \rho_2}{4 k^2}\nonumber \\ &+&\frac{\left(2 \pi ^2
   \left(k-4 \zeta _2\right) \zeta _1 \chi _0\right) \rho_1 \rho_2}{k}+\frac{\left(c^2 \pi ^2 \left(k-4 \zeta _2\right) \zeta _1 \chi
   _2\right) \epsilon _1^2 \rho_1 \rho_2}{k}-\frac{\left(2 c \pi ^2
   \left(k-4 \zeta _2\right) \zeta _1 \chi _1\right) \epsilon _2 \rho_1 \rho_2}{k}\nonumber \\ &+&\frac{\left(c^3 \pi ^2 \left(-11 k+36 \zeta _2\right) \zeta _1 \chi_3\right) \epsilon _1^2 \epsilon _2 \rho_1 \rho_2}{8
   k}+\frac{\left(c^2 \pi ^2 \left(7 k-20 \zeta _2\right) \zeta _1 \chi _2\right)
   \epsilon _2^2 \rho_1 \rho_2}{4 k}\nonumber \\ &-&\frac{\left(c^3 \pi ^2
   \left(k-4 \zeta _2\right) \zeta _1 \chi _3\right) \epsilon _2^3 \rho_1 \rho_2}{4 k}+\left(4 c \pi ^2 \zeta _2 \zeta _1 \chi _1\right) \epsilon _1 \rho_1^2 \rho_2+\frac{1}{16} \left(11 c^3 \pi ^2 \zeta _2 \zeta _1
   \chi _3\right) \epsilon _1^3 \rho_1^2 \rho_2\nonumber \\ &-&\frac{1}{4}
   \left(17 c^2 \pi ^2 \zeta _2 \zeta _1 \chi _2\right) \epsilon _1 \epsilon _2 \rho_1^2 \rho_2+\left(2 c^3 \pi ^2 \zeta _2 \zeta _1 \chi _3\right)
   \epsilon _1 \epsilon _2^2 \rho_1^2 \rho_2-\frac{\left(2 c \pi ^2
   \left(k-3 \zeta _2\right) \zeta _1 \chi _1\right) \epsilon _1 \rho_2^2}{k}\nonumber \\ &+&\frac{\left(c^3 \pi ^2 \left(-5 k+17 \zeta _2\right) \zeta _1 \chi_3\right) \epsilon _1^3 \rho_2^2}{24 k}+\frac{\left(c^2 \pi ^2 \left(k-5
   \zeta _2\right) \zeta _1 \chi _2\right) \epsilon _1 \epsilon _2 \rho_2^2}{2
   k}-\frac{3 c^3 \pi ^2 \left(k-3 \zeta _2\right) \zeta _1 \chi _3}{4 k}\nonumber \\ &-&\left(4 \pi ^2\zeta _2 \zeta _1 \chi _0\right) \rho_1 \rho_2^2-\left(2 c^2 \pi
   ^2 \zeta _2 \zeta _1 \chi _2\right) \epsilon _1^2 \rho_1 \rho_2^2+\left(4 c \pi ^2 \zeta _2 \zeta _1 \chi _1\right) \epsilon _2 \rho_1 \rho_2^2+\frac{1}{4} \left(7 c^3 \pi ^2 \zeta _2 \zeta _1 \chi
   _3\right) \epsilon _1^2 \epsilon _2 \rho_1 \rho_2^2\nonumber \\ &-&\frac{1}{2}\left(3 c^2 \pi ^2 \zeta _2 \zeta _1 \chi _2\right) \epsilon _2^2 \rho_1
   \rho_2^2+\frac{1}{3} \left(2 c^3 \pi ^2 \zeta _2 \zeta _1 \chi _3\right)
   \epsilon _2^3 \rho_1 \rho_2^2+\frac{1}{2} \left(c \pi ^2 \zeta
   _2 \zeta _1 \chi _1\right) \epsilon _1 \rho_2^3+\frac{1}{6} \left(c^3 \pi
   ^2 \zeta _2 \zeta _1 \chi _3\right) \epsilon _1^3 \rho_2^3\nonumber \\ &-&\frac{1}{2}
   \left(3 c^2 \pi ^2 \zeta _2 \zeta _1 \chi _2\right) \epsilon _1 \epsilon _2 \rho_2^3+\frac{1}{8} \left(3 c^3 \pi ^2 \zeta _2 \zeta _1 \chi _3\right)
   \epsilon _1 \epsilon _2^2 \rho_2^3
\label{eq:two_mode_rho2}
\end{eqnarray}
\begin{eqnarray}
   \dot{\rho_2} &=& \frac{\left(4 c^2 \pi ^2 \left(k-\zeta _2\right) \zeta _1 \chi _2\right) \epsilon_1^2}{k^3}+\left(\frac{16 B \pi ^2}{k}-\frac{16 c^2 \pi ^2}{k}-\alpha  (1+\alpha
   )-\frac{c \left(k^3+16 k \pi ^2 \zeta _1 \chi _1 - 16 \pi ^2 \zeta _2 \zeta _1 \chi_1\right)}{k^3}\right) \epsilon _2 \nonumber \\ &-& \frac{\left(4 c^3 \pi ^2 \left(k-\zeta _2\right)
   \zeta _1 \chi _3\right) \epsilon _1^2 \epsilon _2}{k^3}-\frac{\left(2 c^3 \pi ^2
   \left(k-\zeta _2\right) \zeta _1 \chi _3\right) \epsilon _2^3}{k^3}\nonumber \\ &+&\left(4 B \pi ^2-4c^2 \pi ^2-\frac{1}{2} k \alpha  (1+\alpha )-\frac{4 c \pi ^2 \left(3 k-5 \zeta
   _2\right) \zeta _1 \chi _1}{k^2}\right) \epsilon _1 \rho_1-\frac{\left(c^3
   \pi ^2 \left(7 k-11 \zeta _2\right) \zeta _1 \chi _3\right) \epsilon _1^3 \rho_1}{3 k^2}\nonumber \\ &+&\frac{\left(8 c^2 \pi ^2 \left(2 k-3 \zeta _2\right) \zeta _1 \chi_2\right) \epsilon _1 \epsilon _2 \rho_1}{k^2}-\frac{\left(c^3 \pi ^2
   \left(11 k-17 \zeta _2\right) \zeta _1 \chi _3\right) \epsilon _1 \epsilon _2^2 \rho_1}{2 k^2}+\frac{\left(4 \pi ^2 \left(k-4 \zeta _2\right) \zeta _1 \chi
   _0\right) \rho_1^2}{k}\nonumber \\ &+&\frac{\left(c^2 \pi ^2 \left(3 k-10 \zeta _2\right)\zeta _1 \chi _2\right) \epsilon _1^2 \rho_1^2}{k}-\frac{\left(8 c \pi ^2
   \left(k-3 \zeta _2\right) \zeta _1 \chi _1\right) \epsilon _2 \rho_1^2}{k}+\frac{\left(c^3 \pi ^2 \left(-13 k+40 \zeta _2\right) \zeta _1 \chi
   _3\right) \epsilon _1^2 \epsilon _2 \rho_1^2}{4 k}\nonumber \\ &+&\frac{\left(c^2 \pi ^2\left(5 k-16 \zeta _2\right) \zeta _1 \chi _2\right) \epsilon _2^2 \rho_1^2}{2 k}-\frac{\left(c^3 \pi ^2 \left(k-3 \zeta _2\right) \zeta _1 \chi
   _3\right) \epsilon _2^3 \rho_1^2}{k}+\left(6 c \pi ^2 \zeta _2 \zeta _1 \chi
   _1\right) \epsilon _1 \rho_1^3+\frac{1}{24} \left(25 c^3 \pi ^2 \zeta _2
   \zeta _1 \chi _3\right) \epsilon _1^3 \rho_1^3\nonumber \\ &-&\frac{1}{2} \left(13 c^2 \pi^2 \zeta _2 \zeta _1 \chi _2\right) \epsilon _1 \epsilon _2 \rho_1^3+\frac{1}{4} \left(11 c^3 \pi ^2 \zeta _2 \zeta _1 \chi _3\right) \epsilon
   _1 \epsilon _2^2 \rho_1^3+\left(-k-16 d \pi ^2+\frac{16 \pi ^2 \left(k-2
   \zeta _2\right) \zeta _1 \chi _0}{k^2}\right) \rho_2\nonumber \\ &+&\frac{\left(4 c^2 \pi ^2\left(k-2 \zeta _2\right) \zeta _1 \chi _2\right) \epsilon _1^2 \rho_2}{k^2}-\frac{\left(c^3 \pi ^2 \left(5 k-8 \zeta _2\right) \zeta _1 \chi
   _3\right) \epsilon _1^2 \epsilon _2 \rho_2}{k^2}+\frac{\left(2 c^2 \pi ^2
   \left(5 k-8 \zeta _2\right) \zeta _1 \chi _2\right) \epsilon _2^2 \rho_2}{k^2}\nonumber \\ &-&\frac{\left(8 c \pi ^2 \left(k-4 \zeta _2\right) \zeta _1 \chi
   _1\right) \epsilon _1 \rho_1 \rho_2}{k}+\frac{\left(c^3 \pi ^2
   \left(-17 k+62 \zeta _2\right) \zeta _1 \chi _3\right) \epsilon _1^3 \rho_1
   \rho_2}{12 k}\nonumber \\ &+&\frac{\left(3 c^2 \pi ^2 \left(3 k-10 \zeta _2\right) \zeta _1 \chi _2\right) \epsilon _1 \epsilon _2 \rho_1 \rho
   _{\text{b2}}}{k}+\frac{\left(c^3 \pi ^2 \left(-5 k+16 \zeta _2\right) \zeta _1 \chi
   _3\right) \epsilon _1 \epsilon _2^2 \rho_1 \rho_2}{k}-16 \pi ^2
   \zeta _2 \zeta _1 \chi _0 \rho_1^2 \rho_2\nonumber \\ &-&\frac{15}{2} c^2 \pi ^2\zeta _2 \zeta _1 \chi _2 \epsilon _1^2 \rho_1^2 \rho_2+14 c \pi
   ^2 \zeta _2 \zeta _1 \chi _1 \epsilon _2 \rho_1^2 \rho_2+\frac{15}{2} c^3 \pi ^2 \zeta _2 \zeta _1 \chi _3 \epsilon _1^2 \epsilon _2
   \rho_1^2 \rho_2-8 c^2 \pi ^2 \zeta _2 \zeta _1 \chi _2 \epsilon
   _2^2 \rho_1^2 \rho_2\nonumber \\ &+&\frac{13}{6} c^3 \pi ^2 \zeta _2 \zeta _1 \chi _3 \epsilon _2^3 \rho_1^2 \rho_2+\frac{c^2 \pi ^2 \left(2 k-7
   \zeta _2\right) \zeta _1 \chi _2 \epsilon _1^2 \rho_2^2}{k}+\frac{4 c \pi ^2
   \left(-2 k+7 \zeta _2\right) \zeta _1 \chi _1 \epsilon _2 \rho_2^2}{k}+\frac{c^3 \pi ^2 \left(-2 k+7 \zeta _2\right) \zeta _1 \chi _3
   \epsilon _1^2 \epsilon _2 \rho_2^2}{k}\nonumber \\ &+&\frac{\left(c^3 \pi ^2 \left(-4 k+13\zeta _2\right) \zeta _1 \chi _3\right) \epsilon _2^3 \rho_2^2}{3 k}+\left(13
   c \pi ^2 \zeta _2 \zeta _1 \chi _1\right) \epsilon _1 \rho_1 \rho_2^2+\frac{1}{4} \left(9 c^3 \pi ^2 \zeta _2 \zeta _1 \chi _3\right) \epsilon
   _1^3 \rho_1 \rho_2^2\nonumber \\ &-&\left(14 c^2 \pi ^2 \zeta _2 \zeta _1 \chi_2\right) \epsilon _1 \epsilon _2 \rho_1 \rho_2^2+\frac{1}{4}
   \left(25 c^3 \pi ^2 \zeta _2 \zeta _1 \chi _3\right) \epsilon _1 \epsilon _2^2 \rho_1 \rho_2^2-\left(8 \pi ^2 \zeta _2 \zeta _1 \chi _0\right) \rho_2^3\nonumber \\ &-&\left(2 c^2 \pi ^2 \zeta _2 \zeta _1 \chi _2\right) \epsilon _1^2 \rho_2^3+\left(2 c^3 \pi ^2 \zeta _2 \zeta _1 \chi _3\right) \epsilon _1^2
   \epsilon _2 \rho_2^3-\left(4 c^2 \pi ^2 \zeta _2 \zeta _1 \chi _2\right)
   \epsilon _2^2 \rho_2^3
\end{eqnarray}
\section{Solutions from Boundary Layer Analysis}
\label{app:boundlayer}
Here we display the solutions to $\rho$ and $\epsilon$ in the inner region for a right moving pulse, at each order, 

\subsection{Zeroth order}
\begin{eqnarray}
    \epsilon_0 &=& -1 - c_2 \frac{v}{D} \xi \\
    \rho_0 &=& \frac{R \sqrt{\frac{c_2 v^2}{D^2}} e^{-\frac{c_2 \xi ^2 v^2}{2 D^2}}}{\sqrt{2 \pi }}
\end{eqnarray}
\subsection{First order}
\begin{eqnarray}
    \epsilon_1 &=& -\frac{\frac{b c_2 \xi ^2 v^2}{D^2}-\frac{\zeta _1 R \sqrt{\frac{c_2 v^2}{D^2}} \text{Erf}\left(\frac{\sqrt{c_2} \xi  v}{\sqrt{2} D}\right)}{\sqrt{c_2}}}{2 v}-\frac{\zeta _1 \xi  R \sqrt{\frac{c_2 v^2}{D^2}}}{\sqrt{2 \pi } D}
    \label{eq:eps1solnbl}
\end{eqnarray}
\begin{eqnarray}
    \rho_1 &=& \frac{e^{-\frac{c_2 \xi ^2 v^2}{2 D^2}} \left(-\frac{\sqrt{2 \pi } b c_2 \xi ^3 R v^3 \sqrt{\frac{c_2 v^2}{D^2}}}{3 D}-\frac{c_2 \zeta _1 \xi ^2 R^2 v^4}{D^2}+2 \zeta _1 R^2 v^2 e^{-\frac{c_2 \xi ^2 v^2}{2 D^2}}+\frac{\sqrt{2 \pi } \sqrt{c_2} \zeta _1 \xi  R^2 v^3 \text{erf}\left(\frac{\sqrt{c_2} \xi  v}{\sqrt{2} D}\right)}{D}\right)}{4 \pi  D^2 v}\\\nonumber&-&\frac{\zeta _1 R^2 v e^{-\frac{c_2 \xi ^2 v^2}{2 D^2}}}{2 \pi  D^2}
    \label{eq:rho1solnbl}
\end{eqnarray}
\subsection{Second order}
\begin{eqnarray}
    \epsilon_2 &=& -\frac{B^2 c_2 \xi ^3 v}{6 D^3}+\frac{B \zeta _1 \xi  R \sqrt{\frac{c_2 v^2}{D^2}} \text{Erf}\left(\frac{\sqrt{c_2} \xi  v}{\sqrt{2} D}\right)}{2 \sqrt{c_2} D v}+\frac{B \zeta _1 \xi ^2 R \sqrt{\frac{c_2 v^2}{D^2}} e^{-\frac{c_2 \xi ^2 v^2}{2 D^2}}}{6 \sqrt{2 \pi } D^2}-\frac{B \zeta _1 \xi ^2 R \sqrt{\frac{c_2 v^2}{D^2}}}{2 \sqrt{2 \pi } D^2}\\ \nonumber&+&\frac{2 \sqrt{\frac{2}{\pi }} B \zeta _1 R \sqrt{\frac{c_2 v^2}{D^2}} e^{-\frac{c_2 \xi ^2 v^2}{2 D^2}}}{3 c_2 v^2}-\frac{2 \sqrt{\frac{2}{\pi }} B \zeta _1 R \sqrt{\frac{c_2 v^2}{D^2}}}{3 c_2 v^2}-\frac{c_2 \xi ^3 v^3}{6 D^3}-\frac{\zeta _1^2 R^2 e^{-\frac{c_2 \xi ^2 v^2}{2 D^2}} \text{Erf}\left(\frac{\sqrt{c_2} \xi  v}{\sqrt{2} D}\right)}{2 \sqrt{2 \pi } \sqrt{c_2} D^2}+\frac{\zeta _1^2 R^2 \text{Erf}\left(\frac{\sqrt{c_2} \xi  v}{D}\right)}{2 \sqrt{\pi } \sqrt{c_2} D^2}\\ \nonumber &-&\frac{3 \zeta _1^2 R^2 \text{Erf}\left(\frac{\sqrt{c_2} \xi  v}{\sqrt{2} D}\right)}{4 \sqrt{2 \pi } \sqrt{c_2} D^2}+\frac{\zeta _1^2 \xi  R^2 v e^{-\frac{c_2 \xi ^2 v^2}{2 D^2}}}{4 \pi  D^3}-\frac{\xi ^2 v^2}{2 D^2}
\end{eqnarray}
\begin{eqnarray}
    \rho_2  &=& \frac{R e^{-\frac{c_2 \xi ^2 v^2}{2 D^2}}}{24 \pi ^{3/2} D^4 \sqrt{\frac{c_2 v^2}{D^2}}} \Bigg[\frac{\pi  b^2 c_2^3 \xi ^6 v^6}{3 \sqrt{2} D^4}+\frac{\sqrt{2} \pi  b \zeta _1 R \sqrt{\frac{c_2 v^2}{D^2}} \left(3 c_2 D^2 \xi ^2 v^2-c_2^2 \xi ^4 v^4+6 D^4\right)  \text{Erf}\left(\frac{\sqrt{c_2} \xi  v}{\sqrt{2} D}\right)}{\sqrt{c_2} D^2} + 9 \sqrt{2} \zeta _1^2 R^2 v^2 \nonumber \\ &+& \frac{6 \sqrt{2 \pi } \sqrt{c_2} \zeta _1^2 \xi  R^2 v^3 \text{Erf}\left(\frac{\sqrt{c_2} \xi  v}{D}\right)}{D}+\frac{3 \pi  \zeta _1^2 R^2 v^2 \left(\frac{c_2 \xi ^2 v^2}{D^2}-1\right) \text{Erf}\left(\frac{\sqrt{c_2} \xi  v}{\sqrt{2} D}\right){}^2}{\sqrt{2}}\nonumber \\  &-& \frac{3 \sqrt{\pi } \sqrt{c_2} \zeta _1^2 \xi  R^2 v^3 e^{-\frac{c_2 \xi ^2 v^2}{2 D^2}} \left(e^{\frac{c_2 \xi ^2 v^2}{2 D^2}} \left(\frac{c_2 \xi ^2 v^2}{D^2}+5\right)-2\right) \text{Erf}\left(\frac{\sqrt{c_2} \xi  v}{\sqrt{2} D}\right)}{D}\nonumber \\ &-&\zeta _1 R v \left(-e^{-\frac{c_2 \xi ^2 v^2}{D^2}}\right) \biggl(4 \sqrt{\pi } b D \xi  \sqrt{\frac{c_2 v^2}{D^2}} \left(4 e^{\frac{c_2 \xi ^2 v^2}{D^2}}-e^{\frac{c_2 \xi ^2 v^2}{2 D^2}}\right)+9 \sqrt{2} \zeta _1 R v \left(2 e^{\frac{c_2 \xi ^2 v^2}{2 D^2}}-1\right) \nonumber \\ &-&\frac{3 \sqrt{\pi } \sqrt{c_2} \zeta _1^2 \xi  R^2 v^3 e^{-\frac{c_2 \xi ^2 v^2}{2 D^2}} \left(e^{\frac{c_2 \xi ^2 v^2}{2 D^2}} \left(\frac{c_2 \xi ^2 v^2}{D^2}+5\right)-2\right) \text{Erf}\left(\frac{\sqrt{c_2} \xi  v}{\sqrt{2} D}\right)}{D} \nonumber \\ &+& \frac{c_2^2 \xi ^4 v^4 \left(\zeta _1 R v \left(4 \sqrt{\pi } b D \xi  \sqrt{\frac{c_2 v^2}{D^2}}+3 \sqrt{2} \zeta _1 R v\right)-2 \sqrt{2} \pi  D^2 \left(b^2+v^2\right)\right)}{4 D^4} \nonumber \\  &+&\frac{c_2 \xi ^2 v^3 e^{-\frac{c_2 \xi ^2 v^2}{2 D^2}} \left(-2 \sqrt{\pi } b D \zeta _1 \xi  R \sqrt{\frac{c_2 v^2}{D^2}}-\sqrt{2} v e^{\frac{c_2 \xi ^2 v^2}{2 D^2}} \left(2 \pi  D \xi  v-3 \zeta _1^2 R^2\right)-3 \sqrt{2} \zeta _1^2 R^2 v\right)}{D^2}\biggr)\Bigg]
\end{eqnarray}
\section{Numerical Analysis \& Movies}\label{app:numerical}
The PDE system Eqs.\eqref{eq:epsnum_1d} and \eqref{eq:rhonum_1d} is solved using a pseudospectral method in the Dedalus environment. Initial condition is always 
random noise perturbations about an unstrained mesh, $\epsilon = 0$ with homogeneous density $\rho=\rho_0$. For the numerical phase diagrams Fig\,\ref{fig:numphasdiag}, we use 256 modes ($N_x = 256$) in a domain length of $L_x = 10$  with a time step (dt) for the $\alpha=3$ case is $5\times10^{-4}$ and for $\alpha=-2.32$ case is $2\times10^{-3}$ . 
\\
\\
\href{https://drive.google.com/file/d/1K9z5q68VvDUTUv_6kdqjRTvCCc-mUROf/view?usp=sharing}{Movie C1} \label{mov:pulse} Movie shows a travelling pulse solution with parameters $B = 5, \zeta_1 = 20$ with rest of the model parameters as given in Fig.\,\ref{fig:numphasdiag}(a). We see small travelling domains coalesce into a large compact travelling pulse which moves which a constant shape and velocity. 
\\
\\
\href{https://drive.google.com/file/d/1SNJP0RPTqOWwmZjp9dbaX_9em9P4Fubd/view?usp=sharing}{Movie C2} An aperiodic solution at the phase boundary between Segregated and travelling pulse of Fig.\,\ref{fig:numphasdiag}(a) with $B = 2$ and $\zeta_1 = 4.25$ and $L_x = 30$. A similar solution is used to find the Lyapunov Spectrum in Sect. \ref{sect:stchaos}
\\
\\
\href{https://drive.google.com/file/d/1rBMDOCdXO-qWcJ8cUNfmLa2NbdRtX2B-/view?usp=sharing}{Movie C3} A wavetrain solution is seen with $B = 2$ and $\zeta_1 = 17$. Here we see a spatial oscillatory pattern move as a wave with constant velocity. Rest of the parameter values are given in the caption of Fig.\,\ref{fig:numphasdiag}(b). 
\\
\\
\href{https://drive.google.com/file/d/1-i6B-OL5MURs9cJeCB2S3PhU4-PiQVCZ/view?usp=sharing}{Movie C4} A simplified set of equations (Eqs.\,\eqref{eq:simplifiedeps} and \eqref{eq:simplifiedrho} in Sect.\ref{sect:nuctrans}) sufficiently recaptures the travelling pulse solution shown by the full pde. 
\bibliographystyle{unsrt}
\bibliography{apssamp}

\end{document}